%% file: bare_jrnl.tex
\documentclass[journal]{IEEEtran}
\usepackage{graphicx}
\usepackage{amsmath}
\usepackage{amssymb}
\usepackage{color}
\usepackage{multicol, blindtext}
\usepackage{booktabs}
\usepackage{soul}
\usepackage{csquotes}
\usepackage[acronym]{glossaries}
\usepackage{csquotes}
\usepackage{hyperref}
\usepackage{multirow}
\usepackage{caption}

\newcommand\eg{\emph{e.g.,}} 
\newcommand\ie{\emph{i.e.}}

\newcommand\etal{\emph{et al.}}

\definecolor{mypurple}{RGB}{160, 48, 160}
\definecolor{mygreen}{RGB}{84, 130, 53}
\definecolor{myorange}{RGB}{238, 131, 61}
\definecolor{myblue}{RGB}{143, 171, 219}
\definecolor{mybrown}{RGB}{190, 62, 19}
\ifCLASSINFOpdf
\else
\fi

\begin{document}

\title{Underwater Image Restoration via Contrastive Learning and a Real-world Dataset}

\author{Junlin Han, Mehrdad Shoeiby, Tim Malthus, Elizabeth Botha, Janet Anstee, Saeed Anwar, Ran Wei, Mohammad Ali Armin, Hongdong Li, and Lars Petersson
\thanks{This research was funded by the Commonwealth Scientific and Industrial Research Organisation (CSIRO) through its Future Science Platform on Active Integrated Matter under the GBR Test Bed, Workpackages 05 and 11. (\textit{Corresponding author: Junlin Han.)}}
\thanks{Junlin Han, Mehrdad Shoeiby, Saeed Anwar, Ran Wei, Mohammad Ali Armin, and Lars Petersson are with Data61, Commonwealth Scientific and Industrial Research Organisation (CSIRO), Canberra, ACT, 2601, Australia (email: Junlin.Han@data61.csiro.au; Mehrdad.Shoeiby@gmail.com;  Saeed.Anwar@data61.csiro.au; Ran.Wei@data61.csiro.au; Ali.Armin@data61.csiro.au; Lars.Petersson@data61.csiro.au).}
\thanks{Tim Malthus is with Oceans \& Atmosphere, Commonwealth Scientific and Industrial Research Organisation (CSIRO), Brisbane, QLD, 4102, Australia (email: Tim.Malthus@csiro.au).}
\thanks{Elizabeth Botha and Janet Anstee are with Oceans \& Atmosphere, Commonwealth Scientific and Industrial Research Organisation (CSIRO), Canberra, ACT, 2601, Australia (email: Elizabeth.Botha@csiro.au; Janet.Anstee@csiro.au).}
\thanks{Hongdong Li is with College of
Engineering \& Computer Science, Australian National University (ANU), Canberra, ACT, 2601, Australia (email: honddong.li@anu.edu.au).}}

%
%

\markboth{Journal of \LaTeX\ Class Files,~Vol.~13, No.~9, September~2014}%
{Shell \MakeLowercase{\textit{et al.}}: Bare Demo of IEEEtran.cls for Journals}
%



\maketitle

\begin{abstract}
Underwater image restoration is of significant importance in unveiling the underwater world. Numerous techniques and algorithms have been developed in the past decades. However, due to fundamental difficulties associated with imaging/sensing, lighting, and refractive geometric distortions, in capturing clear underwater images, no comprehensive evaluations have been conducted of underwater image restoration. To address this gap, we have constructed a large-scale real underwater image dataset, dubbed `HICRD' (Heron Island Coral Reef Dataset), for the purpose of benchmarking existing methods and supporting the development of new deep-learning based methods. We employ accurate water parameter (diffuse attenuation coefficient) in generating reference images. There are 2000 reference restored images and 6003 original underwater images in the unpaired training set. Further, we present a novel method for underwater image restoration based on unsupervised image-to-image translation framework. Our proposed method leveraged contrastive learning and generative adversarial networks to maximize the mutual information between raw and restored images. Extensive experiments with comparisons to recent approaches further demonstrate the superiority of our proposed method. Our code and dataset are publicly available at \href{https://github.com/JunlinHan/CWR}{\textcolor{red}{GitHub}}.
\end{abstract}



%
\IEEEpeerreviewmaketitle

\section{Introduction}
\label{sec:introduction}
%
%
%
%
\IEEEPARstart{F}{or} marine science and ocean engineering, significant applications such as the surveillance of coral reefs, underwater robotic, and submarine cable inspection require clear underwater images. For high-level computer vision tasks~\cite{reggiannini2021use}, clear underwater images are also required. Furthermore, clear images help the development of undersea remote sensing techniques~\cite{williams2014exploiting,ludeno2018microwave,fei2014contributions}. Raw images with low visual quality do not meet these expectations, where the clarity of raw images is degraded by both absorption and scattering~\cite{carlevaris2010initial,Akkaynak_2018_CVPR,yuan2020underwater}. The clarity of underwater images thus plays an essential role in scientific missions. Therefore, fast, accurate, and effective techniques producing clear underwater images need to be developed to improve the visibility, contrast, and color properties of underwater images for satisfactory visual quality and practical usage. The main techniques to infer clear images are known as underwater image enhancement and restoration. Underwater image enhancement aims to produce visually pleasing results, focusing on enhancing contrast and brightness. In comparison, underwater image restoration is based on the image formulation model~\cite{carlevaris2010initial,Akkaynak_2018_CVPR,yuan2020underwater}, aims to rectify the distorted colors and present the true colors of the underwater scene. In this paper, we mainly focus on the objective of underwater image restoration.
\begin{figure*}[!htb]
     \includegraphics[width=18 cm ]
     {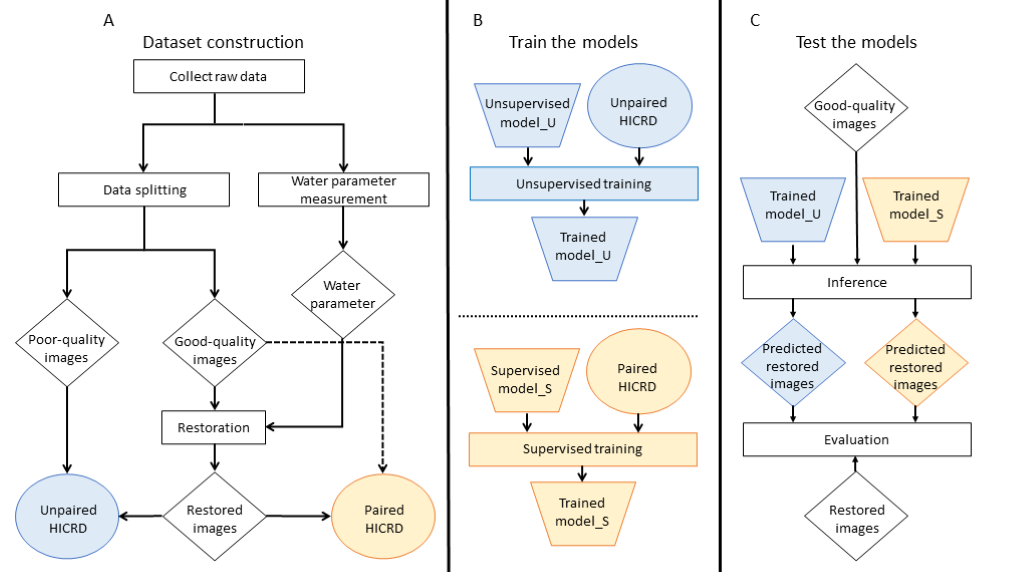}
     \caption{The flowchart of this paper, where rectangles denote operations, rhombus denote images, circles denote datasets, and trapezoids denote models. We show (A) Dataset Construction (Section~\ref{sec:dataset}): We construct a large-scale dataset, referred to as Heron Island Coral Reef Dataset (HICRD). We collect raw data from different sites, split the raw images into poor-quality and good-quality images. We restore the good-quality images with the measured water parameters. Paired HICRD contains the good-quality images, and the corresponding reference restored images, while unpaired HICRD contains poor-quality images and reference restored images. (B) Train the models (Section~\ref{sec:method} and Section~\ref{sec:experiment}): We mainly train the unsupervised models using the unpaired HICRD. Paired HICRD can support the training of supervised models. (C) Test the models (Section~\ref{sec:experiment}): We test the trained models using the same test set, measure the predicted restored images with reference restored images.}
     \label{fig:flow}
\end{figure*}

In the underwater scene, visual quality is greatly affected by light refraction, absorption, and scattering. For instance, underwater images usually have a green-bluish tone since red light with longer wavelengths attenuates faster. Underwater image restoration is a highly ill-posed problem, which requires several parameters (\eg{} global background light and medium transmission map) that are mostly unavailable in practice.

These parameters can be roughly estimated by employing priors and supplementary information; we refer to these approaches as the conventional methods. Most conventional underwater restoration methods employ priors to estimate the unknown parameters in the imaging model. The Dark Channel Prior (DCP)~\cite{he2010single} is a simple yet effective prior for image dehazing. With the same imaging model, some variants based on the DCP have been developed for underwater imaging~\cite{carlevaris2010initial,drews2013transmission,galdran2015automatic}. Carlevaris \etal{}~\cite{carlevaris2010initial} exploited the differences between attenuation among three color channels as a prior to predict the transmission map of the underwater scene. Drews \etal{}~\cite{drews2013transmission} ignored the red channel and proposed the Underwater Dark Channel Prior (UDCP). The UDCP applies DCP to the blue and green channels only to predict the transmission map. The Red channel prior~\cite{galdran2015automatic} makes use of the red channel, \ie{}, the color associated with the longest wavelengths, to restore the underwater image. 

Chiang \etal{}~\cite{chiang2011underwater} employed measured attenuation coefficients of open ocean waters and assumed that the transmission map of the red channel is the recovered transmission by which the underwater image can be restored.
Differing from~\cite{drews2013transmission}, which focued on the green and blue channels, Lu \etal{}~\cite{lu2015contrast} ignored the green channel and restored the underwater image by leveraging the transmission map estimated from the blue and red channels only. 
Peng \etal{}~\cite{peng2017underwater} proposed a method based on image blurriness and light absorption (IBLA), which uses a depth estimation approach to restore underwater images.
Relying on the Jerlov water types~\cite{jerlov1976marine}, Berman \etal{}~\cite{berman2020underwaterpami} took spectral profiles of water types into account to simplify underwater image restoration as image dehazing. However, due to the diversity of water types and lighting conditions, conventional underwater image restoration methods can fail to rectify the color of underwater images. 

Another approach to underwater image restoration employs learning-based methods. Recent advances in deep learning demonstrate dramatic success in different fields. In the line of learning-based underwater image restoration, Cao \etal{}~\cite{cao2018underwaterestorationr} recovered the scene radiance based on estimated background light and scene depth using two neural networks.
Barbosa \etal{}~\cite{barbosa2018visualrestoration} employed the DehazeNet~\cite{cai2016dehazenet} to predict the transmission map, thus predicting the scene radiance.
The Underwater residual convolutional neural network (URCNN)~\cite{hou2018jointrestoration} is a residual convolutional neural network. It learns the transmission map to restore the image. After restoration, it also enhances the restored image. 
The Underwater convolutional neural network (UWCNN)~\cite{li2020underwater} is a supervised model including three densely connected building blocks~\cite{huang2017densely} trained on its synthetic underwater image  datasets.

Though learning-based methods have demonstrated the effectiveness in modeling the underwater image restoration process, they require a large-scale dataset for training, which is often difficult to obtain. Thus, most learning-based models, regardless of enhancement or restoration, use either a limited number of real underwater images~\cite{li2019underwater, berman2020underwaterpami, duarte2016dataset}, synthesized underwater images~\cite{li2020underwater,fabbri2018enhancing,islam2020fast,cao2018underwaterestorationr,barbosa2018visualrestoration,hou2018jointrestoration,wang2019underwater}, or natural in-air images~\cite{li2018emerging,wang2019underwater} as either the source domain or target domain of the training set. Four commonly used underwater image restoration datasets are as follows:

\begin{itemize}
\item \textbf{SQUID\footnote {\href{http://csms.haifa.ac.il/profiles/tTreibitz/datasets/ambient_forwardlooking/index.html}{SQUID}}~\cite{berman2020underwaterpami}:} The Stereo Quantitative Underwater Image Dataset includes 57 stereo pairs from four different sites, two in the Red Sea and the other two in the Mediterranean Sea. Every image contains multiple color charts and its range map without providing the reference images. 

\item \textbf{TURBID\footnote{\href{http://amandaduarte.com.br/turbid/}{TURBID }}~\cite{duarte2016dataset}:} TURBID consists of five different subsets of degraded images with its respective ground-truth reference image. Three subsets are publicly available; they are degraded by milk, deepblue, and chlorophyll. Each subset contains 20, 20, and 42 images, respectively.

\item \textbf{UWCNN Synthetic Dataset\footnote{\href{https://li-chongyi.github.io/proj_underwater_image_synthesis.html}{UWCNN synthetic}}~\cite{li2020underwater}:} UWCNN synthetic dataset contains ten subsets, each subset represents one water type with 5000 training images and 2495 validation images. The dataset is synthesized from the NYU-v2 RGB-D dataset~\cite{Silberman:ECCV12NYU}, The first 1000 clean images are used to generate the training set and the remaining 449 clean images are used to generate the validation set. Each clean image is used to generate five images based on different levels of atmospheric light and water depth.

\item \textbf{Sea-thru dataset\footnote{\href{http://csms.haifa.ac.il/profiles/tTreibitz/datasets/sea_thru/index.html}{Sea-thru}}~\cite{Akkaynak_2019_CVPR}:} This dataset contains five subsets, representing five diving locations. It contains 1157 images in total; every image is with its range map. Color charts are available within the partial dataset. No reference images are provided. 

\end{itemize}

The above datasets are limited in capturing the natural variability in a wide range of water types. The synthetic training data also limits the generalization of models~\cite{anwar2020survey}. Also, only partial datasets provide reference restored images. Lacking a large-scale, publicly available real-world underwater image restoration dataset with scientifically restored images limits the development of learning-based underwater image restoration methods.

To overcome the earlier discussed challenges, we construct a large-scale real-world underwater image dataset, the Heron Island Coral Reef Dataset (HICRD). HICRD contains 9676 raw underwater images in total and 2000 scientifically restored reference images. It contains two subsets, unpaired HICRD and paired HICRD. It also contains the measured diffuse attenuation coefficient and the camera sensor response. 

To fully use the proposed HICRD, we design a learning-based underwater image restoration model trained on HICRD. We formulate the restoration problem as an image-to-image translation problem and propose a novel \textbf{C}ontrastive Under\textbf{W}ater \textbf{R}estoration approach (CWR). CWR combines both contrastive learning~\cite{he2020momentum,chen2020simple} and generative adversarial networks~\cite{goodfellow2014generative}. Given an underwater image as input, CWR directly outputs a restored image, showing the real color of underwater objects as if the image was taken in-air without any structure or content loss. The flowchart of this paper is presented in figure~\ref{fig:flow}.

The main contribution of our work is summarized as follows:
\begin{itemize}
    \item We construct a large-scale, high-resolution underwater image dataset with real underwater images and restored images. The Heron Island Coral Reef Dataset (HICRD) provides a platform to evaluate the performance of various underwater image restoration models on real underwater images with various water types. It also enables the training of both supervised and unsupervised underwater image restoration models.
    
    \item We propose an unsupervised learning-based model, \ie{}, CWR, which leverages contrastive learning to maximize the mutual information between corresponding patches of the raw image and the restored image to capture the content and color feature correspondences between two image domains.
\end{itemize}
Some results of this paper were published originally in its conference version~\cite{han2021cwr}. However, this longer article includes the details of HICRD and CWR to provide a deeper understanding, \eg{} data collection (Section~\ref{sec:data collection}), water parameter measurement (Section~\ref{sec:water parameter}), and implementation details of CWR (Section~\ref{sec:implementation}). We also add more related work in the introduction (Section~\ref{sec:introduction}). A more comprehensive experiment is available in the result section (Section~\ref{sec:experiment}), including more baselines (Section~\ref{sec:baselines}) and more evaluation results (Section~\ref{sec:results}). A discussion section (Section~\ref{sec:discussion}) is added to provide a comprehensive analysis.


\section{A Real-world Dataset}
\label{sec:dataset}
To address the previously discussed issues and gaps, we collect a real-world dataset, measure the water parameters, and create reference restored images. In this section, we elaborate on the constructed dataset in detail, including data collection, water parameter measurement, imaging model, and reference image generation. The flowchart of this section is shown in figure~\ref{fig:flow} A.
\begin{figure*}[!htb]
     \includegraphics[width=18 cm]
     {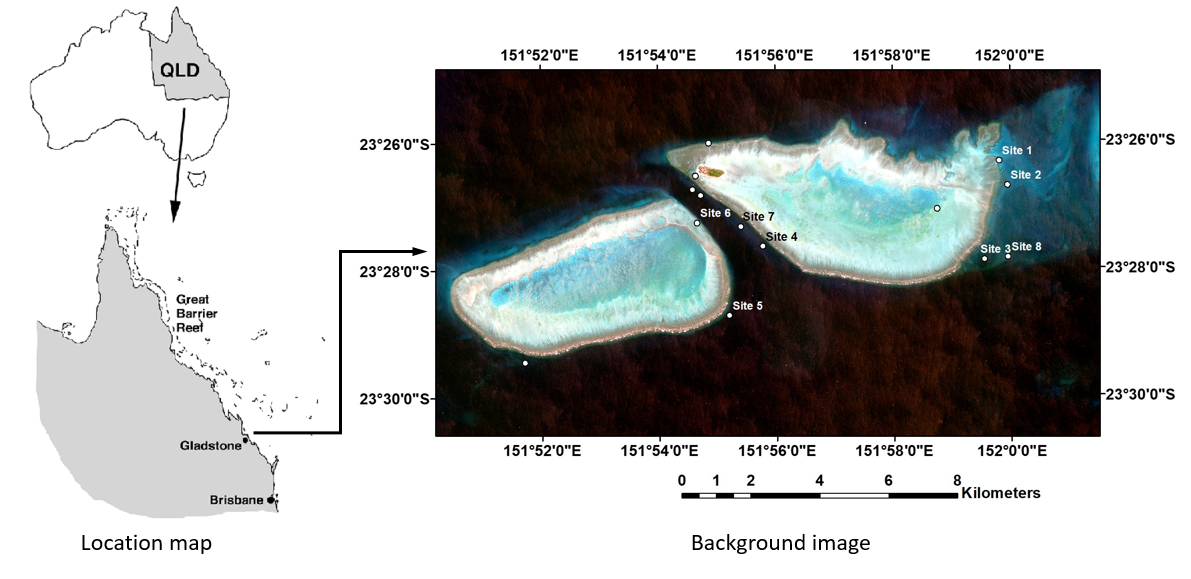}
     \caption{Location of field sites during the 2018 field campaign around Heron and Wistari Reefs. Location map is modified from \cite{schonberg2007bioeroding}. Background image: Sentinel-2 true color image, acquired on 14/05/2021.}
     \label{fig:location}
\end{figure*}

\subsection{Data Collection}
\label{sec:data collection}

Images and other forms of optical data of HICRD were collected from several sample locations around Heron Reef, located in the Southern Great Barrier Reef, Australia, from 11 to 16 June 2018. These sample locations, corresponding to Reef Check Australia permanent transect sites~\cite{reef_check2018}, were selected to represent a variety of coral reef environments with a range of coral densities, water depths, and water conditions. Figure~\ref{fig:location} shows the location of sample stations on Heron Reef and nearby Wistari Reef.

Image acquisition of HICRD was conducted using CSIRO's Underwater Spectral Instrument Platform (CORYCAEUS). RGB images in HICRD were captured by the on board MQ022CG-CM color camera. It is a small form factor USB 3.0 camera made by Ximea GmbH, Germany. The camera carries a 2/3" 10-bit CMOSIS CMV 2000 sensor. The optics is an Edmund Optics 12mm focal length F1.8 lens. This lens has a field of view of 41.1$^{\circ}$ when used with the Ximea camera. During image capturing procedures, the camera aperture was fixed, and its exposure time was adjusted by a human operator on board the boat viewing the live video streamed back from the cameras while divers were swimming underwater. The image capturing frequency was 3 to 5 frames/second.  

Our data acquisition task was accomplished by a team of divers and boat members. The divers carried out underwater data collection using CORYCAEUS. Boat staff made surface measurements, provided safety monitoring, and maintained boat control. The detailed procedure is as follows: On arrival at a coral reef sample site, first, boat crew verify the GPS location, and collect water samples, surface reflectance, water column spectral absorbance and backscattering measurements. CORYCAEUS is turned on and carried to the divers. Diver one first measures underwater reflectance. Diver two then manipulates CORYCAEUS to swim in a fixed direction keeping a constant distance of 2m above the reef. Diver one safeguards diver 2 during the whole measuring transect. Meanwhile, one of the boat members monitors live video from CORYCAEUS transferred by a tethering cable. The boat crew keep the boat following the divers by motoring at a low speed, and keeping a safe distance of (10 to 30 meters) from the divers. At the end of each transect, the boat crew signal the divers to the surface. Both divers and CORYCAEUS are retrieved from the water. Final readings of the GPS position and other measurements are recorded.

\subsection{Water Parameter Estimation}
\label{sec:water parameter}
To understand the effects of bio-optical constituents on the behavior of light through the water column, data was collected to characterize the inherent and apparent optical properties (IOPs and AOPs).
\input{figures/waterparameterdisplay}

At each dive site, IOPs were collected as vertical profiles of the absorption a($\lambda$), and attenuation spectral coefficients c($\lambda$) were measured using a 10cm path length WET Labs (\href{http://seabird.com}{link}) ac-s spectral absorption-attenuation meter. Backscattering coefficients bb($\lambda$) were measured using a WET Labs BB-9 spectral backscattering meter, while temperature, salinity, and density were collected using a WET Labs Water Quality Monitor (WQM). Profiles were measured with all instruments connected to a WET Labs DH-4 data logger, allowing consistent time stamping. The backscattering measurements were corrected for salinity and light loss due to absorption over the path length using the absorption and scattering values from the ac-s~\cite{boss_twardowski}.

Surface water samples were collected for laboratory analysis of the absorbing components (particulate and dissolved). Samples for particulate absorption (phytoplankton and non-algal) were prepared by filtration through a 25mm Whatman GF/F glass-fibre filter and stored flat in liquid nitrogen for transport. Particle absorption coefficients were measured over the 250–800 nm spectral range with a Cintra 4040 UV/VIS dual-beam spectrophotometer (\href{http://gbcsci.com }{link}) equipped with an integrating sphere~\cite{https://doi.org/10.1002/2014JC010205}. Samples for measurement of chromophoric dissolved organic matter (CDOM) absorption were prepared by vacuum filtration through a 0.22 µm Millipore filter and preserved with sodium azide for sample transport~\cite{mannino_novak}. The absorbance of each filtrate was measured from 250 to 800 nm in a 10 cm pathlength quartz cell using a Cintra 4040 UV/VIS spectrophotometer~\cite{https://doi.org/10.1002/2014JC010205}. Total absorption is considered as the sum of the particulate and dissolved components plus that of pure water.

Radiometric profiles (AOP) were measured at each station using three TriOS Ramses radiometers (\href{http://trios.de}{link}). One irradiance radiometer was mounted on the boat to measure total downwelling irradiance, and two sensors were mounted on an underwater frame to measure water radiance and irradiance, respectively. The diffuse attenuation coefficient ($K_{d}$) is determined from measurements of the rate of change (slope) of the logarithm of the irradiance (E) depth profile~\cite{Austin1981} and can be written as:

\begin{equation}
    K_{d} = \frac{1}{z-z_{0}}ln[\frac{E(z)}{E(z_{0})}],
\end{equation}

where z$_0$ is the uppermost depth of measurement. Depth averaged $K_{d}$ over the depth interval from the surface to depth $z$ is expressed as~\cite{Simon:13}:

\begin{equation}
    K_{d}(0 \to z) = \frac{\int_{0}^{z} K_{d}(z)dz}{z}.
\end{equation}

\subsection{Detailed information of HICRD}
The Heron Island Coral Reef Dataset (HICRD) contains raw underwater images from eight sites with detailed metadata for each site, including water parameters (diffuse attenuation), maximum dive depth, and the camera model. Six sites have wavelength-dependent attenuation within the water column. According to depth information of raw images and the distance between objects and the camera, images with roughly the same depth, constant distance, and good visual quality are labeled as good-quality. Images with sharp depth changes, distance changes, or poor visual quality are labeled as low-quality. We apply our imaging model described in section~\ref{sec:imaging} to good-quality images, producing corresponding reference restored images, and manually remove some restored images with non-satisfactory quality.

\begin{figure*}[!htb]
     \centering
     \includegraphics[width=18 cm]
     {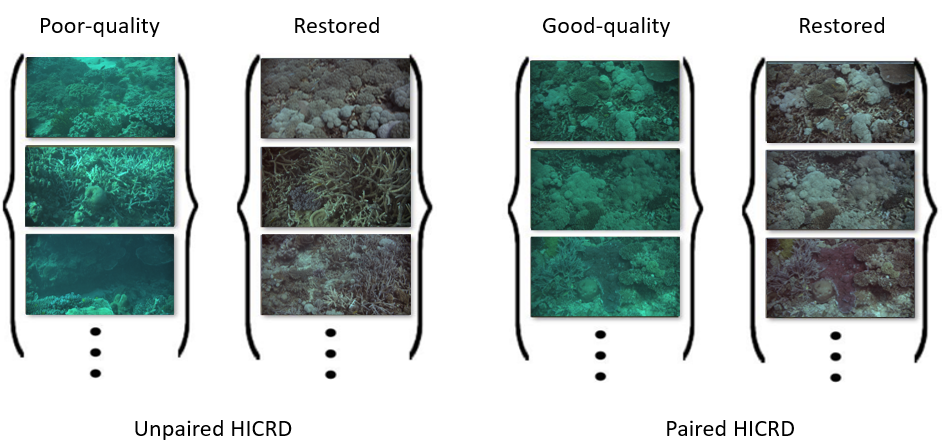}
     \caption{Example images and split of HICRD dataset. Unpaired HICRD contains poor-quality images and restored images as the training set while paired HICRD uses the good-quality images and corresponding reference restored images. }
     \label{fig:dataset}
\end{figure*}

HICRD contains 6003 low-quality images, 3673 good-quality images, and 2000 reference restored images. We use low-quality images and restored images as the unpaired training set. In contrast, the paired training set contains 1700 good-quality images and corresponding restored images. The test set contains 300 good-quality images as well as 300 corresponding restored images as reference images. All images are in 1842 x 980 resolution. Figure~\ref{fig:parameter} shows the total diffuse attenuation for different water types.  Figure~\ref{fig:dataset} shows randomly selected image examples from HICRD and the split of HICRD. Table~\ref{tab:dataset} presents the detailed information of the HICRD dataset. 
\input{tables/dataset}

\subsection{Underwater Imaging Model and Reference Image Generation}
\label{sec:imaging}
Unlike the dehazing model~\cite{he2010single}, absorption plays a critical role in an underwater scenario. Each channel's absorption coefficient is wavelength-dependent, being the highest for red and the lowest for blue. A simplified underwater imaging model~\cite{serikawa2014underwater} can be formulated as:
\begin{equation}
    I^{c}(x) = J^{c}(x)t^{c}(x) + A^{c}(1-t^{c}(x)),~~~~c\in \{{r,g,b}\},
    \label{eq:haze}
\end{equation}
where $I(x)$ is the observed intensity, $J(x)$ is the scene radiance, and $A$ is the global atmospheric light. $t^{c}(x) = e^{-\beta^{c} d(x)}$ is the medium transmission describing $A(x)$ the portion of light that is not scattered, $\beta^{c}$ is the light attenuation coefficient and $d(x)$ is the distance between camera and object. Channels are in RGB space. We assume the measured diffuse attenuation coefficient $K_{d}^{c}$ to be identical to the attenuation coefficient $\beta^{c}$.

\begin{figure}[!htb]
     \centering
     \includegraphics[width=8 cm]
     {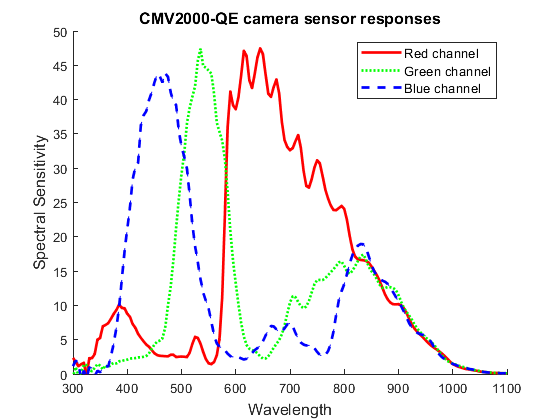}
     \caption{Camera sensor response for camera sensor type CMV2000-QE which is used in collecting real underwater images.}
     \label{fig:response}
\end{figure}

Transmittance is highly related to $\beta^{c}$, which is the light attenuation coefficient of each channel, and it is wavelength-dependent. Unlike previous work~\cite{peng2017underwater,chiang2011underwater}, instead of assigning a fixed wavelength for each channel containing bias (\eg, 600nm, 525nm, and 475nm for red, green, and blue), we employ the camera sensor response to conduct a more accurate estimation. Figure~\ref{fig:response} shows the camera sensor response of camera sensor type CMV2000-QE used in collecting the dataset.

The new total attenuation coefficient is estimated by:
\begin{equation}
    p^{c}=\int_{a}^{b}\beta^{\lambda}S^{c}(\lambda)d\lambda,
    \label{eq:response}
\end{equation}
where $p^{c}$ is the total attenuation coefﬁcient, $\beta^{\lambda}$ is the attenuation coefficient of each wavelength, and $S^{c}(\lambda)$ is the camera sensor response of each wavelength. Following the human visible spectrum, we set a = 400nm and b = 750nm to calculate medium transmission for each channel. We modify medium transmission $t^{c}(x)$ in Equation~\ref{eq:haze} leading to a more accurate estimation: 
\begin{equation}
t^{c}(x) = e^{-p^{c} d(x)}.
    \label{eq:total}
\end{equation}

It is challenging to measure the scene's actual distance from an individual image without a depth map. Instead of using a flawed estimation approach, we assume the distance between the scene and the camera to be small (1m - 5m) and manually assign a distance for each good-quality image. 

The global atmospheric light, $A^{c}$, is usually assumed to be the pixel's intensity with the highest brightness value in each channel. However, this assumption often fails due to the presence of artificial lighting and self-luminous aquatic creatures. Since we have access to the diving depth, we can define $A^{c}$ as follows:
\begin{equation}
    A^{c} = e^{-p^{c}z},
    \label{eq:airlight}
\end{equation}
where $p^{c}$ is the total attenuation coefﬁcient, $z$ is the diving depth.

With the medium transmission and global atmospheric light, we can recover the scene radiance. The final scene radiance $J(x)$ is estimated as:
\begin{equation}
    J^{c}(x) = \frac{I_{c}(x)-A_{c}}{max(t_{c}(x),t_{0})}+A_{c}.
    \label{radiance}
\end{equation}
Typically, we choose $t_{0}$ = 0.1 as a lower bound. In practice, due to the complexity with image formulation, our imaging model based on the simplified underwater imaging model may encounter information loss, \ie, the pixel intensity values of $J^{c}(x)$ are larger than 255 or less than 0. This problem is avoided by only mapping a selected range (13 to 255) of pixel intensity values from $I$ to $J$. However, outliers may still occur; we re-scale the whole pixel intensity values to enhance contrast and keep information lossless after restoration.

\section{Proposed Method}
\label{sec:method}
\begin{figure*}[!htb]
     \includegraphics[width=18 cm ]
     {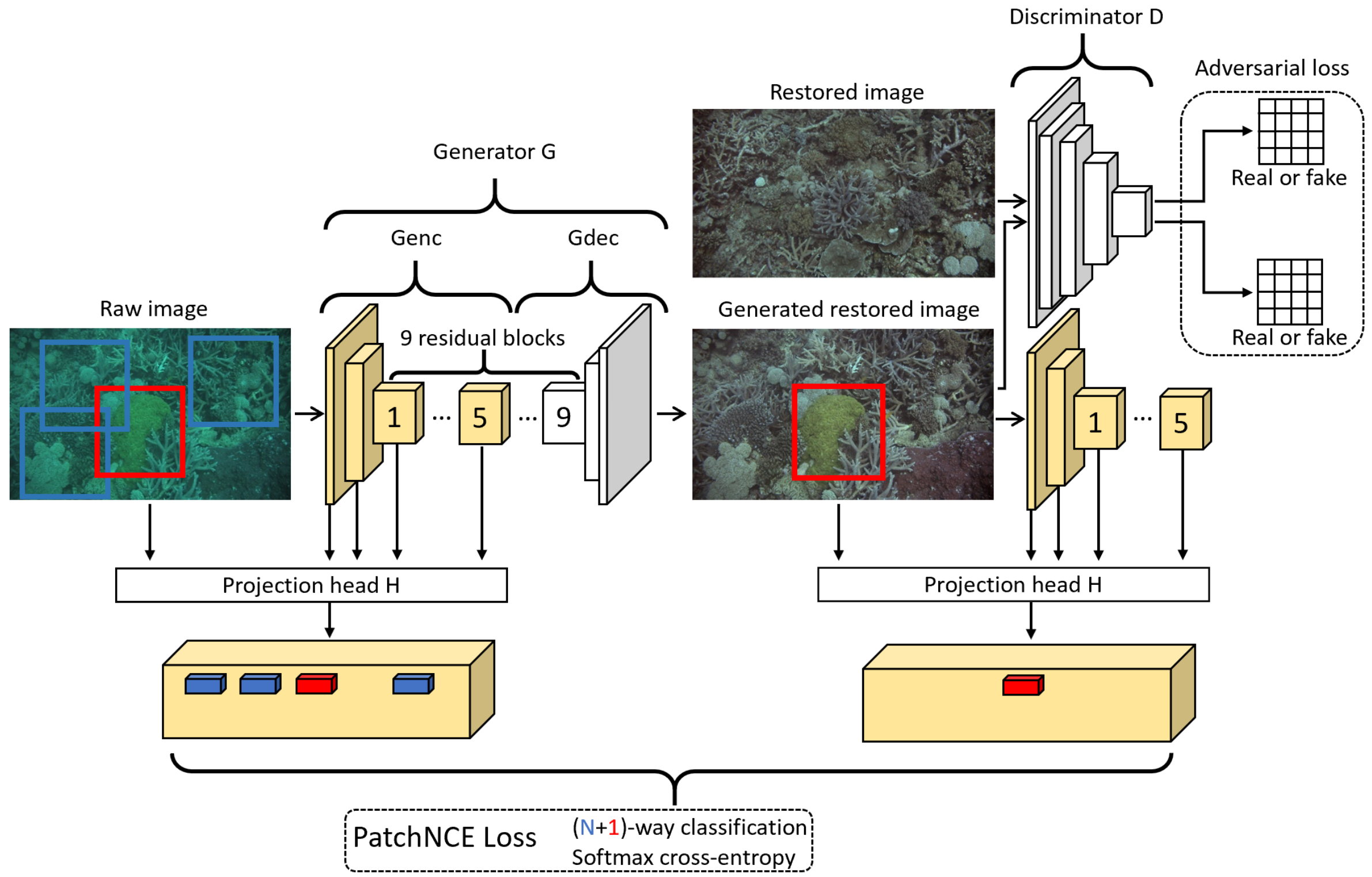}
    \caption{The overall architecture and losses of CWR. CWR targets learning a mapping $G: X\rightarrow Y$, \ie{}, raw underwater image $\rightarrow$ restored image. We use a ResNet-based generator with nine residual blocks and define the first half of the Generator $G$ to be Encoder. We use encoder $G_{enc}$ and projection head $H$ to extract the patch-based, multi-layer features from the raw image and its translated version (generated restored image), embedding one image to a stack of features. Each layer represents a different resolution of the patches. Here, we depict one layer only for simplification. In the stack of features, given the \textcolor{red}{red ``query''} from the generated restored image, we set up an $(N+1)$-way classification problem and compute the probability that \textcolor{red}{a red ``positive''} is selected over \textcolor{blue}{N blue ``negatives''}. Such a process shows the computation of the PatchNCE loss.
    We send both the generated restored image and the accurate restored image to the PatchGAN discriminator for computing the adversarial loss. For each image, the discriminator outputs a result metric showing the discrimination result in the patch level. Note that the patches here are different from that computing the PatchNCE loss. We omit the identity loss here.}
     \label{fig:cwr}
\end{figure*}
Given two domains $\mathcal{X} \subset \mathbb{R}^{H \times W \times 3}$ and $\mathcal{Y} \subset \mathbb{R}^{H \times W \times 3}$ and a dataset of unpaired instances $X$ containing raw underwater images $x$ and $Y$ containing restored images $y$. We denote it $X= \left\{x \in \mathcal{X} \right\}$ and $Y= \left\{y \in \mathcal{Y} \right\}$. We aim to learn a mapping $G : X\rightarrow Y$ to enable underwater image restoration.

\textbf{C}ontrastive Under\textbf{W}ater \textbf{R}estoration (CWR) has a generator $G$ as well as a discriminator $D$. $G$ enables the mapping from domain $X$ to domain $Y$ and $D$ ensures that the translated images belong to the correct image domain. The first half of the generator is defined as an encoder, while the second half is a decoder and denoted $G_{enc}$ and $G_{dec}$ respectively.

For the mapping, we extract features of images from several layers of the encoder and send them to a two-layer MLP projection head. Such a projection head learns to project the extracted features from the encoder to a stack of features. CWR combines three losses, including Adversarial loss, PatchNCE loss, and Identity loss. Figure~\ref{fig:cwr} shows the overall architecture and losses of CWR. The details of our objective are described below.

\subsection{Adversarial Loss}
An adversarial loss is employed to encourage the generator to generate visually similar images to images from the target domain, for the mapping $G : X\rightarrow Y$ with discriminator $D$ , the GAN (Generative Adversarial Network) loss~\cite{goodfellow2014generative} is calculated by:
\begin{equation}
\begin{aligned}
\mathcal{L}_{\mathrm{GAN}}\left(G, D, X, Y\right) &=\mathbb{E}_{y\sim{Y}}\left[\log D(y)\right] \\
&+\mathbb{E}_{x\sim{X}}\left[\log \left(1-D(G(x))\right]\right.,
\end{aligned}
\end{equation}
where $G$ tries to generate images $G(x)$ that look similar to
images from domain $Y$, while $D$ aims to distinguish between translated samples $G(x)$ and real samples $y$.

\subsection{PatchNCE Loss}
Our goal is to maximize the mutual information between corresponding patches of the input and the output. For instance, for a patch showing a coral reef of the generated restored image, we should be able to associate it more strongly to the same coral reef patch of the raw input underwater image other than the rest of the patches of the image. 

Following the setting of CUT~\cite{park2020contrastive}, we use a noisy contrastive estimation framework to maximize the mutual information between inputs and outputs. The idea behind contrastive learning is to correlate two signals, i.e., the ``query'' and its ``positive'' example, in contrast to other examples in the dataset (referred to as ``negatives'').

We map query, positive, and $N$ negatives to $K$-dimensional vectors and denote them $v, v^{+} \in R^{K}$ and $ v^{-} \in R^{N \times K}$ respectively. Note that $v_{n}^{-} \in R^{K}$ denotes the n-th negative. We set up an $(N+1)$-way classification problem and compute the probability that a ``positive'' is selected over ``negatives''. Mathematically, this can be expressed as a cross-entropy loss which is computed by:
\begin{equation}
\begin{aligned}
    &\ell\left(\boldsymbol{v}, \boldsymbol{v}^{+}, \boldsymbol{v}^{-}\right)= \\
    & -\log(\frac{\exp \left( \operatorname{sim}({v},\boldsymbol{v}^{+}) / \tau\right)}{\exp \left( \operatorname{sim}({v},\boldsymbol{v}^{+}) / \tau\right)+\sum_{n=1}^{N} \exp \left( \operatorname{sim}({v},\boldsymbol{v}_{n}^{-}) / \tau\right)}),
\end{aligned}
\label{eq:nce}
\end{equation}
where $\operatorname{sim}(\boldsymbol{u}, \boldsymbol{v})=\boldsymbol{u}^{\top} \boldsymbol{v} /\|\boldsymbol{u}\|\|\boldsymbol{v}\|$ denotes the cosine similarity between $\boldsymbol{u}$ and $\boldsymbol{v}$. $\tau$ denotes a temperature parameter to scale the distance between the query and other examples, we use 0.07 as default. We set the numbers of negatives 255 as the default setting.

We use $G_{enc}$ and a 2-layer projection head $H$ to extract features from domain $X$.
We select $L$ layers from $G_{enc}(X)$ and send them to $H$, embedding one image to a stack of features $\left\{\boldsymbol{z}_{l}\right\}_{L}=\left\{H^{l}\left(G_{\mathrm{enc}}^{l}(\boldsymbol{x})\right)\right\}_{L}$, where $G_{\mathrm{enc}}^{l}$ represents the output of $l$-th selected layers. For the patches, after having a stack of features, each feature actually represents one patch from the image. We take advantage of that and denote the spatial locations in each selected layer as $s \in \{1,...,S_{l}\}$, where $S_{l}$ is the number of spatial locations in each layer.
We select a query each time, referring to the corresponding feature (``positive'') as $\boldsymbol{z}_{l}^{s} \in \mathbb{R}^{C_{l}}$ and all other features (``negatives'') as $\boldsymbol{z}_{l}^{S \backslash s} \in \mathbb{R}^{\left(S_{l}-1\right) \times C_{l}}$, where $C_{l}$ is the number of channels in each layer. 
We aim to match the corresponding patches of input and output images. The patch-based, multi-layer PatchNCE loss for mapping $G : X\rightarrow Y$ can be expressed as:
\begin{equation}
\mathcal{L}_{\mathrm{PatchNCE}}(G,H,X)=
\mathbb{E}_{\boldsymbol{x} \sim X} \sum_{l=1}^{L} \sum_{s=1}^{S_{l}} \ell\left(\hat{z}_{l}^{s}, \boldsymbol{z}_{l}^{s}, \boldsymbol{z}_{l}^{S \backslash s}\right).
\label{eq:patchnce}
\end{equation}

\subsection{Identity Loss}
In order to prevent generators from unnecessary changes and keep the structure identical after translation, we add an identity loss~\cite{CycleGAN2017}.
\begin{equation}
    \mathcal{L}_{\text {Identity}}(G)=\mathbb{E}_{y \sim {Y} }\left[\|G(y)-y\|_{1}\right].
\end{equation}
Such an identity loss also encourages the mappings to preserve color composition between the input and output.

\subsection{Full Objective}
The generated restored image should be realistic ($\mathcal{L}_{GAN}$), and patches in the corresponding input raw and generated restored images should share a correspondence ($\mathcal{L}_{\mathrm{PatchNCE}}$). The generated restored image should have an identical structure to the input raw image. In contrast, the colors are the true colors of scenes ($\mathcal{L}_{\mathrm{Identity}}$). The full objective is:
\begin{equation}
\begin{aligned}
\mathcal{L}(G,D,H)
&=\lambda_{GAN}\mathcal{L}_{GAN}(G,D,X)\\ 
&+\lambda_{NCE}\mathcal{L}_{\mathrm{PatchNCE}}(G, H, X)\\
&+\lambda_{idt}\mathcal{L}_{\text {Identity}}(G).
\end{aligned}
\end{equation}
We set $\lambda_{GAN}$ = 1, $\lambda_{NCE}$ = 1 and $\lambda_{idt}$ = 10.

\subsection{Implementation Details}
\label{sec:implementation}

\subsubsection{Architecture of Generator and Layers Used for PatchNCE Loss}
Figure~\ref{fig:cwr} represents the architecture of the generator and layers used for computing PatchNCE loss and Table~\ref{tab:generator} presents the detailed architecture of the generator. Our generator architecture is based on CycleGAN~\cite{CycleGAN2017} and CUT~\cite{park2020contrastive}. We use a ResNet-based~\cite{he2016deep} generator with nine residual blocks for training. It contains two downsampling blocks, nine residual blocks, and two upsampling blocks. Each downsampling and upsampling block follows two-stride convolution/deconvolution, normalization and an activation function,\ie{}, ReLU. Each residual block contains convolution, normalization, ReLU, convolution, normalization, and a residual connection.
\input{tables/generator}

We define the first half of generator $G$ as an encoder represented as $G_{enc}$. The patch-based multi-layer PatchNCE loss is computed using features from five layers of the encoder (the RGB pixels, the first and second downsampling convolution, and the first and the fifth residual block). The patch sizes extracted from these four layers are 1x1, 9$\times$9, 15$\times$15, 35$\times$35, and 99$\times$99 resolution, respectively. Following CUT~\cite{park2020contrastive}, for each layer's features, we sample 256 random locations and apply a 2-layer MLP (projection head $H_{X}$) to infer 256-dim final features.

\subsubsection{Architecture of Discriminator}
We use the same PatchGAN discriminator architecture as CycleGAN~\cite{CycleGAN2017} and Pix2Pix~\cite{isola2017image}, which uses local patches of sizes 70x70 and assigns a result to every patch. This is equivalent to manually cropping one image into 70x70 overlapping patches, running a regular discriminator over each patch, and averaging the results. The PatchGAN discriminator is shown in Figure~\ref{fig:cwr}. For instance, the discriminator takes an image from either domain $X$ or domain $Y$, passes it through five downsampling Convolutional-Normalization-LeakeyReLU layers, and outputs a result matrix of 62x62. Each element corresponds to the classification result of one patch. The detailed architecture of discriminator is presented in Table~\ref{tab:discriminator}. 
\input{tables/discriminator}

\section{Results}
\label{sec:experiment}
\subsection{Baselines}
\label{sec:baselines}
Most learning-based underwater image restoration methods do not provide the source code, and do not support unsupervised training; instead, we train a set of state-of-the-art image-to-image translation models, using the HICRD dataset to enable underwater image restoration. We focus on unsupervised image-to-image translation models to fully exploit HICRD.
We compare CWR to several state-of-the-art baselines from different views, including image-to-image translation approaches (CUT~\cite{park2020contrastive}, CycleGAN~\cite{CycleGAN2017} and DCLGAN~\cite{han2021dcl}), conventional underwater image enhancement methods (Histogram-prior~\cite{li2016hist}, Retinex~\cite{fu2014retinex} and Fusion~\cite{ancuti2017color}), conventional underwater image restoration methods (UDCP~\cite{drews2013transmission}, DCP~\cite{he2010single}, IBLA~\cite{peng2017underwater}, and Haze-line~\cite{berman2020underwaterpami}), and learning-based restoration method (UWCNN~\cite{li2020underwater}). We use the pre-trained UWCNN model with water type-3, which is close to our dataset. 

For image-to-image translation approaches, CUT~\cite{park2020contrastive} and DCLGAN~\cite{han2021dcl} aim to maximize the mutual information between corresponding patches of the input and the output. DCLGAN~\cite{han2021dcl} employs a dual learning setting and assigns different encoders to different domains, to gain a better performance. We firstly employ CUT and DCLGAN to enable the underwater image restoration task. CycleGAN~\cite{CycleGAN2017} is based on the cycle-consistency assumption, \ie{}, it learns the reverse mapping from the target domain back to the source domain and forces the reconstruction image to be identical to the input image. CycleGAN has been widely used in the field of underwater image restoration and enhancement. We treat those image-to-image translation approaches trained on HICRD as learning-based underwater image restoration methods.

For conventional underwater image enhancement methods, Histogram-prior~\cite{li2016hist} is based on a histogram distribution prior; it contains two steps, where the first step is an underwater image dehazing, and the second step is a contrast enhancement. Retinex~\cite{fu2014retinex} enhances a single underwater image with a color correction step, a layer decomposition step, and a color enhancement step. Fusion~\cite{ancuti2017color} creates a color-compensated version and a white-balanced version based on the input underwater image. Using two different versions of images derived from the input underwater image as inputs; it employs a multi-scale fusion step to enable underwater image enhancement.

Other restoration methods are introduced in section~\ref{sec:introduction}.

\subsection{Training Details}
We train CWR, CUT, CycleGAN, and DCLGAN for 100 epochs with the same learning rate of 0.0002. The learning rate decays linearly after half the epochs. We load all images in 800$\times$800 resolution and randomly crop them into 512$\times$512 patches during training. We load test images in 1680$\times$892 resolution for all methods. 
For CWR, we employ spectral normalization~\cite{miyato2018spectral} for the discriminator and instance normalization~\cite{ulyanov2016instance} for the generator. The batch size is set to one. The ADAM~\cite{kingma2014adam} optimizer is employed for optimization, we set $\beta_{1}$ = 0.5 and $\beta_{2}$ = 0.999. We train our method and other baselines using a Tesla P100-PCIE-16GB GPU. The GPU driver version is 440.64.00, and the CUDA version is 10.2.

\input{figures/result_long}
\subsection{Evaluation Protocol}
\label{sec:evaluation}
To fully measure the performance of different methods, we employed three full-reference metrics: Mean-Square Error (MSE), Peak signal-to-noise ratio (PSNR), and structural similarity index (SSIM)~\cite{ssim2014} as well as a non-reference metric designed for underwater images: Underwater Image Quality Measure (UIQM)~\cite{panetta2015human}. UIQM
comprises three underwater image attribute measures: the underwater image colorfulness measure (UICM), the underwater image sharpness measure (UISM), and the underwater image contrast measure (UIConM). A higher UIQM score suggests that the result is more consistent with human visual perception.

We additionally use Fréchet Inception Distance (FID)~\cite{TTUR} to measure the quality of generated images. FID embeds a set of generated samples into the feature space given by a particular layer of InceptionNet, treating the embedding layer as a continuous multivariate Gaussian that estimates the mean and covariance of the generated data and the real data. The Fréchet distance between these two Gaussian curves is then used to quantify the quality of the generated samples. Mathematically, it can be written as:
\begin{equation}
    FID(r, g)=\left\|\mu_{r}-\mu_{g}\right\|_{2}^{2}+T r\left(\Sigma_{r}+\Sigma_{g}-2\left(\Sigma_{r} \Sigma_{g}\right)^{\frac{1}{2}}\right).
\end{equation}
where $\mu_{r},\mu_{g}$ are the means of real data and generated data respectively, $\Sigma_{r} ,\Sigma_{g}$ are the covariances of real data and generated data.
A lower FID score suggests generated images tend to be more realistic.

For all metrics, we use the full test set for evaluation, \ie{}, 300 underwater raw images, and 300 corresponding restored images. We also compare the evaluation speed for all methods.
\input{tables/compare}

\subsection{Evaluation Results}
\label{sec:results}
Table~\ref{tab:3compare} provides a quantitative evaluation, where no method always wins in terms of all metrics. However, CWR performs stronger than all the baseline in full-reference metrics. CWR also shows competitive results in UIQM and FID score. This suggests that CWR performs accurate restoration, and the outputs of CWR are consistent with human visual perception. Restoration methods tend to have a lower FID score, while enhancement methods tend to have a higher UIQM score. Restoration methods provide more similar outputs to our reference restored images, while the outputs of enhancement methods are more consistent with human visual perception. This is due to the objectives of restoration and enhancement being different. Conventional restoration methods run slowly, while enhancement methods tend to run faster. 

Figure~\ref{fig:result1} presents the randomly selected qualitative results. Learning-based methods provide better restoration results than conventional restoration methods, where conventional restoration methods fail to perform the correct restoration. Such a phenomenon is shown in UDCP~\cite{drews2013transmission}, DCP~\cite{he2010single}, and Haze-line~\cite{berman2020underwaterpami}, where they fail to remove the green-bluish tone in underwater images. CWR performs a better restoration process than other learning-based methods in keeping with the structure and content of the restored images identical to raw images with negligible artifacts.

\input{figures/result_enhancement2}

For underwater image enhancement methods, Figure~\ref{fig:result2} shows the qualitative results of Example 2 and Example 3 with zoomed patches. Conventional enhancement methods keep the structure unchanged, but sometimes over-enhance the images, adding a bright color to the objects.

\section{Discussion}
\label{sec:discussion}
This paper described a large-scale real-world underwater restoration dataset, offering high-quality training data and proper reference restored images to support the development of both unsupervised and supervised learning-based models. Using the HICRD dataset, we evaluated recent methods from different views, including image-to-image translation approaches, conventional underwater image enhancement methods, conventional underwater image restoration methods, and learning-based restoration methods. In addition, a novel unsupervised method leveraging state-of-the-art contrastive learning techniques is proposed to fully capitalize on the HICRD dataset. In quantitative evaluations, though CWR performs better overall, there is no one method that always wins in both full-reference and non-reference metrics, suggesting new learning-based underwater image restoration methods should be developed.

Though HICRD is a large-scale dataset, it does not cover all of the common water types. All images are acquired from Heron Island, where different sites still share similar environmental and geomorphological conditions. Most images within HICRD are related to coral reefs. HICRD did not capture other objects present underwater, such as shipwrecks, underwater archaeological sites~\cite{mangeruga2018guidelines}.

We simply assume the attenuation diffuse attenuation $K_{d}$ and the (horizontal) attenuation coefficient $\beta$ to be identical. Such an assumption is widely used in our community~\cite{berman2020underwaterpami,li2020underwater,berman2017diving}; however, it is an incorrect assumption~\cite{Akkaynak_2017_CVPR,Akkaynak_2018_CVPR}. $K_{d}$ is an Apparent optical property (AOP); it varies with sun-sensor geometry while $\beta$ is an inherent optical property (IOP). This assumption leads to an imprecise restoration process; thus, an imperfect reference restored image. Moreover, following the commonly used simplified underwater imaging model consequently lead to errors. The simplified underwater imaging model ignores the absorption coefficient and ignores the difference between backward scattering and attenuation, where attenuation is the combination of scattering and absorption. Recently, a revised underwater imaging model~\cite{Akkaynak_2018_CVPR} was proposed. However, due to the complexity of this revised underwater imaging model, only a few works followed this model. More complexity leads to more parameters to estimate, where estimates inevitably contain more errors compared to accurate measurements.
We do not employ the revised underwater imaging model as our imaging model to mitigate the errors from the estimation of parameters. New underwater image restoration datasets should be constructed using a more precise manner, that is, employing the revised underwater imaging with fewer assumptions to generate more precise reference restored images. 

The evaluation metrics for underwater image restoration are limited. Existing non-reference methods are mainly designed for underwater image enhancement; thus, if no reference restored images are provided, it would be difficult to evaluate the restoration methods. Using underwater image enhancement metrics to evaluate the restoration methods may lead to failure cases~\cite{anwar2020survey}. New evaluation metrics that incorporate underwater physical model properties should be developed to advance the underwater image restoration research. Also, more specialized high-level task-driven evaluation metrics should be developed. Liu \etal{}~\cite{liu2020real} showed that the image quality assessment and detection accuracy are not strongly correlated. However, one goal of underwater image enhancement and restoration is to produce clear images to support applications in practice. Novel task-driven evaluation metrics should be developed, and they should correlate with the quality of input images.

In this paper, we focus on RGB underwater image restoration. Multi-spectral and hyper-spectral images can be used for various underwater applications ~\cite{yi2018instantaneous,fu2020underwaterhyper,mogstad2019shallow,dumke2018underwater}. However, few multi-spectral and hyper-spectral underwater image restoration techniques have been developed~\cite{guo2016model}. New work in underwater image restoration can consider building a benchmark dataset for multi-spectral or hyper-spectral images to further research along this critical direction.

\section{Conclusion}
This paper presents a real-world underwater image dataset HICRD that offers large-scale data consisting of real underwater images and restored images enabling a comprehensive evaluation of existing underwater image enhancement \& restoration methods. HICRD employs measured water parameters to create the reference restored images. We believe that HICRD will enable a significant advancement of the use of learning-based underwater restoration methods, in both unsupervised and supervised manners. Along with HICRD, a novel approach, CWR, employing contrastive learning, is proposed exploiting HICRD. CWR performs a realistic underwater image restoration process in an unsupervised manner. Experimental results show that CWR performs significantly better than several conventional enhancement and restoration methods while showing more desirable results than state-of-the-art learning-based restoration methods.


%



\section*{Acknowledgment}
The authors would like to thank Dr. Phillip Ford (Ocean \& Atmosphere, CSIRO) for editing and reviewing the manuscript. The authors would also like to thank CBRMPA (Great Barrier Reef Marine Park Authority) for permitting us to conduct experiments at the Great Barrier Reef (Permit G18/40267.I). The authors would also like to thank the anonymous reviewers for their reviews.

\ifCLASSOPTIONcaptionsoff
  \newpage
\fi



%

\bibliographystyle{IEEEtran}
\bibliography{mybib}

%








\end{document}

%% file: figures/waterparameterdisplay.tex
\begin{figure*}[!htbp]
  \begin{minipage}[t]{0.49\linewidth} 
    \centering 
    \includegraphics[scale = 0.61]{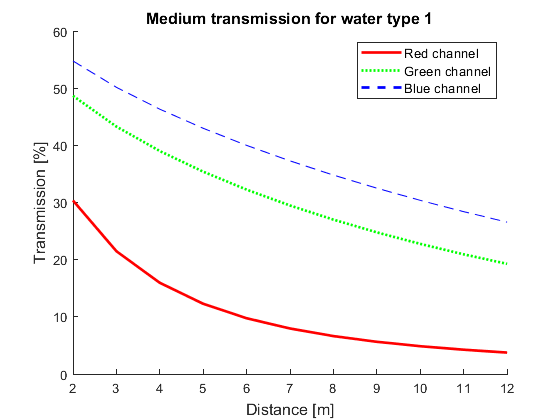}
  \end{minipage} 
    \begin{minipage}[t]{0.49\linewidth} 
    \centering 
    \includegraphics[scale = 0.61]{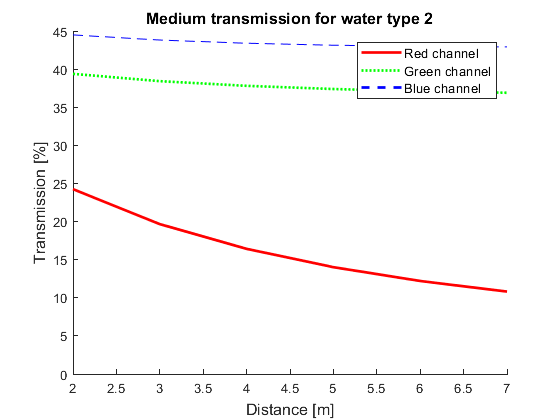}
  \end{minipage} 
  \\
  \begin{minipage}[t]{0.49\linewidth} 
    \centering 
    \includegraphics[scale = 0.61]{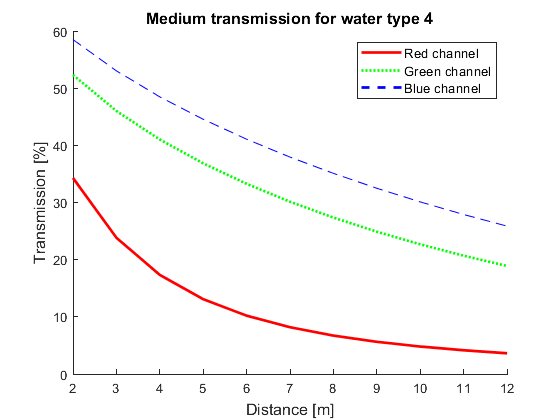}
  \end{minipage} 
    \begin{minipage}[t]{0.49\linewidth} 
    \centering 
    \includegraphics[scale = 0.61]{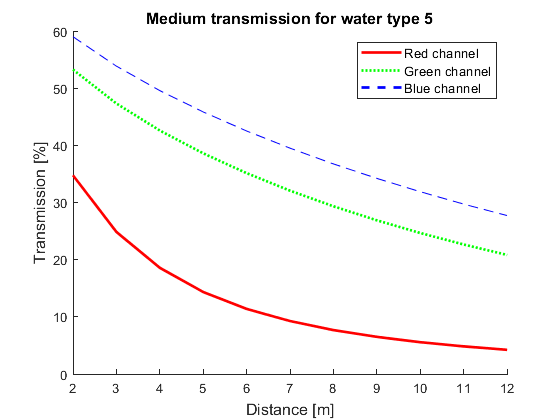}
  \end{minipage} 
  \\
    \begin{minipage}[t]{0.49\linewidth} 
    \centering 
    \includegraphics[scale = 0.61]{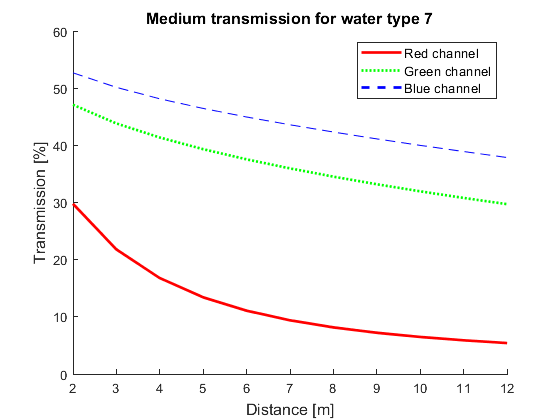}
  \end{minipage} 
    \begin{minipage}[t]{0.49\linewidth} 
    \centering 
    \includegraphics[scale = 0.61]{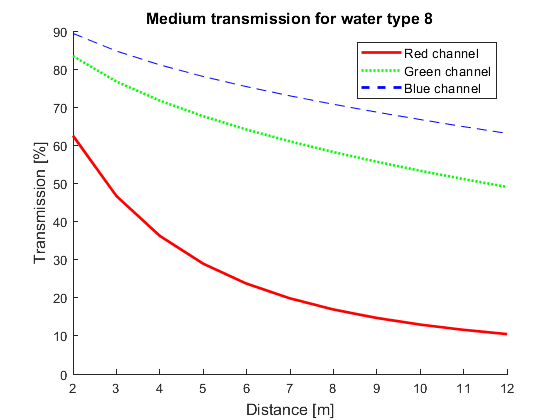}
  \end{minipage} 
  
  \caption{The medium transmission $t^{c}(x)$ for six different sites/water types. HICRD contains various water types, where type1, type2, type 4, type 7, and type 8 are different. The medium transmission is formulated in equation~\ref{eq:total}.}
  \label{fig:parameter}
\end{figure*}

%% file: tables/dataset.tex
\begin{table*}[!htbp]
  \centering
  \fontsize{11.5}{15}\selectfont
\begin{tabular}{ccccccccc}
\toprule
       &Site 1 & Site 2 &Site 3&Site 4&Site 5&Site 6 & Site 7  & Site 8\\
\midrule 
Low-quality images   & 47 &53 &458 &1923&1726 & 117 &961&718      \\
Good-quality images  & 43 &71 &151 &1042&1344&52 &677&293         \\
Reference images & 25 &16 &N.A.&160&1241&N.A. &457& 101     \\
Water type    & 01 &02 &N.A.&04  &05&N.A.&07 & 08    \\
Diver 1 max depth &6.8&5.7&7.3&8.7&7.2&6.6&10.7&9.4  \\  
Diver 2 max depth &6.4&6.3&7.6&8.8&7.4&6.5&10.4&9.3   \\
\bottomrule
\end{tabular}
\caption{Detailed information of HICRD dataset. Unpaired training set contains all low-quality images and all reference restored images while the paired training set contains 1700 good-quality images and corresponding reference restored images. Test set contains 300 good-quality images from site 5, as well as 300 corresponding reference restored images. The water parameter for Site 3 and Site 6 are not available.}
\label{tab:dataset}
\end{table*}

%% file: tables/generator.tex
\begin{table*}[!htbp] 
\centering 
\fontsize{9}{15}\selectfont
{\begin{tabular}{lcccc} 
\hline 
\multicolumn{2}{c}{\multirow{2}*{\text{Component}}} &\multirow{2}*{\text{Input $\rightarrow$ Output}} & \text{\multirow{2}*{Layer information}}\\ 
& & shape & \\
\hline 
\multirow{8}{*}{Encoder}
& conv layer& (3,512,512) $\rightarrow$ (64,512,512) & ReflectionPad, Conv(3,64,7,1), IN, ReLU\\ 
& \textcolor{blue}{\textbf{\textit{downsampling 1}} } & (64,512,512) $\rightarrow$ (128,256,256)& Conv(64,128,3,2),IN,ReLU \\ 
& \textcolor{blue}{\textbf{\textit{downsampling 2}} } & (128,256,256) $\rightarrow$ (256,128,128)& Conv(128,256,3,2),IN,ReLU \\ 
&\textcolor{blue}{\textbf{\textit{residual block 1}} }  & (256,128,128) $\rightarrow$ (256,128,128) &Conv(256,256,3,1),IN,ReLU,Conv(256,256,3,1),IN, RC \\ 
& residual block 2 & (256,128,128) $\rightarrow$ (256,128,128) & Conv(256,256,3,1),IN,ReLU,Conv(256,256,3,1),IN, RC \\ 
& residual block 3 & (256,128,128) $\rightarrow$ (256,128,128) & Conv(256,256,3,1),IN,ReLU,Conv(256,256,3,1),IN, RC \\ 
& residual block 4 & (256,128,128) $\rightarrow$ (256,128,128) & Conv(256,256,3,1),IN,ReLU,Conv(256,256,3,1),IN, RC \\ 
&\textcolor{blue}{\textbf{\textit{residual block 5}} }& (256,128,128) $\rightarrow$ (256,128,128) &Conv(256,256,3,1),IN,ReLU,Conv(256,256,3,1),IN, RC  \\ 
\hline 
\multirow{6}{*}{Decoder} 
& residual block 6 & (256,128,128) $\rightarrow$ (256,128,128) & Conv(256,256,3,1),IN,ReLU,Conv(256,256,3,1),IN, RC  \\ 
& residual block 7 & (256,128,128) $\rightarrow$ (256,128,128) & Conv(256,256,3,1),IN,ReLU,Conv(256,256,3,1),IN, RC \\ 
& residual block 8 & (256,128,128) $\rightarrow$ (256,128,128) &Conv(256,256,3,1),IN,ReLU,Conv(256,256,3,1),IN, RC  \\ 
& residual block 9 & (256,128,128) $\rightarrow$ (256,128,128) & Conv(256,256,3,1),IN,ReLU,Conv(256,256,3,1),IN, RC \\ 
& upsampling 1 & (256,128,128) $\rightarrow$ (128,256,256)& DeConv(I256,O128,K3,S2),SN,ReLU \\ 
& upsampling 2 & (128,256,256) $\rightarrow$ (64,512,512)& DeConv(I128,O256,K3,S2),SN,ReLU \\ 
& conv layer& (64,512,512) $\rightarrow$ (3,512,512) & ReflectionPad, Conv(I64,O3,K7,S1), SN, Tanh\\ 
\hline 
\end{tabular}} 
\caption{Architecture of Generator. IN means instance normalization~\cite{ulyanov2016instance} and RC denotes residual connection~\cite{he2016deep}. For a Conv(3,64,7,1), it refers to input channels = 3, output channels = 64, kernel size = 7, and stride = 1, respectively. The layers used for PatchNCE Loss are in bold-italic format with blue color.} 
 \label{tab:generator}
\end{table*}

%% file: tables/discriminator.tex
\begin{table*}[!htbp] 
\centering 
\fontsize{9}{15}\selectfont
{\begin{tabular}{lcccc} 
\hline 
\multicolumn{2}{c}{\multirow{2}*{\text{Component}}} &\multirow{2}*{\text{Input $\rightarrow$ Output}} & \text{\multirow{2}*{Layer information}}\\ 
& & shape & \\
\hline 
\multirow{5}{*}{Discriminator}
& downsampling 1 & (3,512,512) $\rightarrow$ (64,256,256)& Conv(3,64,4,2),SN,LReLU \\ 
& downsampling 2 & (64,256,256) $\rightarrow$ (128,128,128)& Conv(64,128,4,2),SN,LReLU \\ 
& downsampling 3 & (128,128,128) $\rightarrow$ (256,64,64)& Conv(128,256,4,2),SN,LReLU \\ 
& downsampling 4 & (256,64,64) $\rightarrow$ (512,63,63)& Conv(128,256,4,1),SN,LReLU \\ 
& downsampling 5 & (512,63,63) $\rightarrow$ (1,62,62)& Conv(512,1,4,1)\\ 
\hline 
\end{tabular}} 
\caption{Architecture of PatchGAN discriminator. SN means spectral normalization~\cite{miyato2018spectral} and LReLU denotes LeakyReLU, we use negative slope = 0.2. For a 2D convolution with input channels = 3, output channels = 64, kernel size = 4, and stride = 2 is referred to as Conv(3,64,4,2). The output of discriminator is a 62 x 62 result metric.} 
 \label{tab:discriminator}
\end{table*}

%% file: figures/result_long.tex
\begin{figure*}[!htbp]

  \begin{minipage}[t]{0.02\linewidth} 
  \text{}
  \end{minipage} 
    \begin{minipage}[t]{0.19\linewidth} 
    \centering 
    \text{Example 1}
  \end{minipage}   
  \begin{minipage}[t]{0.19\linewidth} 
    \centering 
    \text{Example 2}
  \end{minipage}   
  \begin{minipage}[t]{0.19\linewidth} 
    \centering 
    \text{Example 3}
  \end{minipage}   
  \begin{minipage}[t]{0.19\linewidth} 
    \centering 
    \text{Example 4}
  \end{minipage} 
    \begin{minipage}[t]{0.19\linewidth} 
    \centering 
    \text{Example 5}
  \end{minipage} 
\\
    \begin{minipage}[t]{0.02\linewidth} 
    \centering
    \rotatebox{90}{ \quad  \small Input}
          \vspace{-20mm}
      \end{minipage} 
     \begin{minipage}[t]{0.19\linewidth} 
    \centering 
    \includegraphics[width=1.32in, height=0.72in]{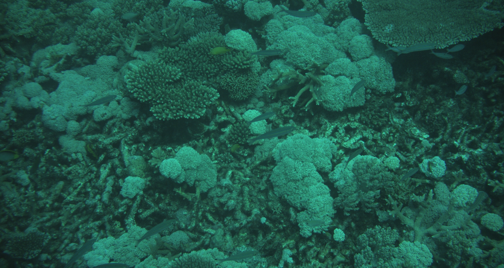}
  \end{minipage} 
       \begin{minipage}[t]{0.19\linewidth} 
    \centering 
    \includegraphics[width=1.32in, height=0.72in]{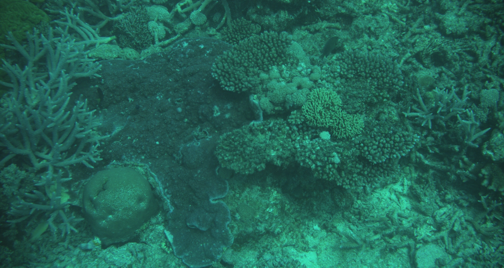}
  \end{minipage} 
       \begin{minipage}[t]{0.19\linewidth} 
    \centering 
    \includegraphics[width=1.32in, height=0.72in]{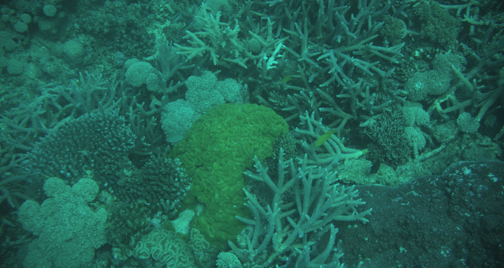}
  \end{minipage} 
       \begin{minipage}[t]{0.19\linewidth} 
    \centering 
    \includegraphics[width=1.32in, height=0.72in]{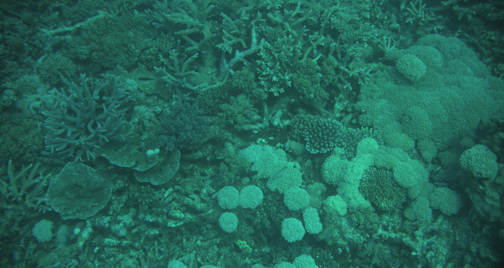}
  \end{minipage} 
       \begin{minipage}[t]{0.19\linewidth} 
    \centering 
    \includegraphics[width=1.32in, height=0.72in]{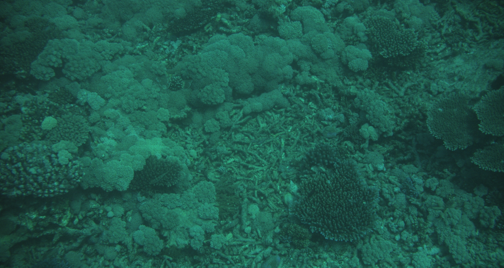}
  \end{minipage}   
  \\
\begin{minipage}[t]{0.02\linewidth} 
    \centering
    \rotatebox{90}{ \; \small UDCP}
              \vspace{-200mm}
      \end{minipage} 
     \begin{minipage}[t]{0.19\linewidth} 
    \centering 
    \includegraphics[width=1.32in, height=0.72in]{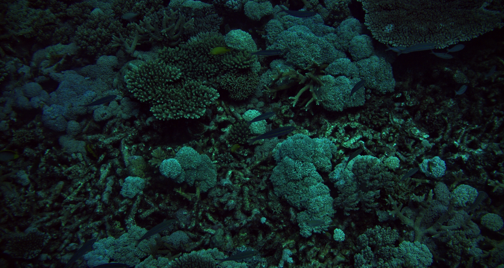}
  \end{minipage} 
       \begin{minipage}[t]{0.19\linewidth} 
    \centering 
    \includegraphics[width=1.32in, height=0.72in]{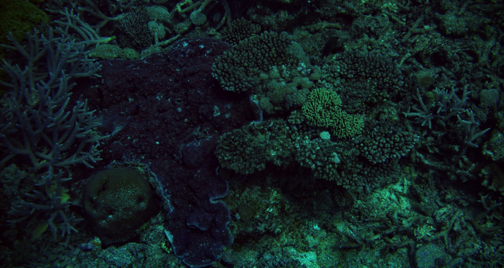}
  \end{minipage} 
       \begin{minipage}[t]{0.19\linewidth} 
    \centering 
    \includegraphics[width=1.32in, height=0.72in]{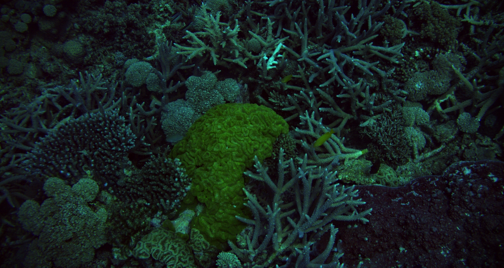}
  \end{minipage} 
       \begin{minipage}[t]{0.19\linewidth} 
    \centering 
    \includegraphics[width=1.32in, height=0.72in]{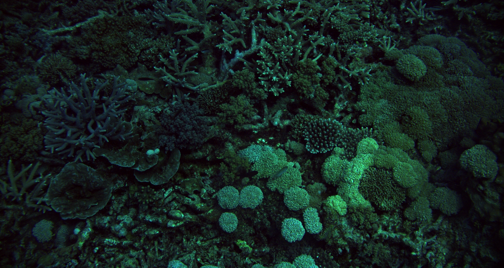}
  \end{minipage} 
         \begin{minipage}[t]{0.19\linewidth} 
    \centering 
    \includegraphics[width=1.32in, height=0.72in]{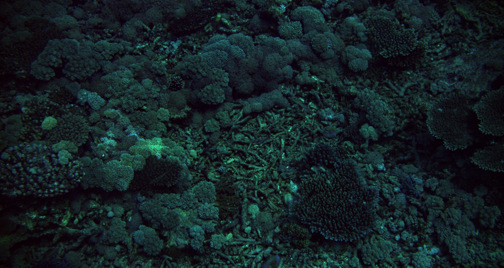}
  \end{minipage} 
  \\   
\begin{minipage}[t]{0.02\linewidth} 
    \centering
    \rotatebox{90}{ \;\; \;  \small DCP}
              \vspace{-200mm}
      \end{minipage} 
     \begin{minipage}[t]{0.19\linewidth} 
    \centering 
    \includegraphics[width=1.32in, height=0.72in]{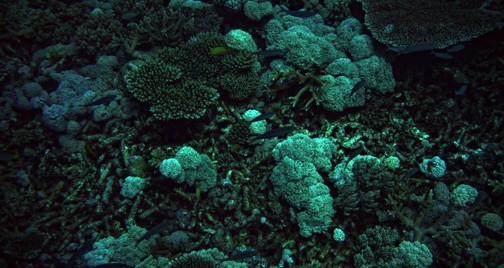}
  \end{minipage} 
       \begin{minipage}[t]{0.19\linewidth} 
    \centering 
    \includegraphics[width=1.32in, height=0.72in]{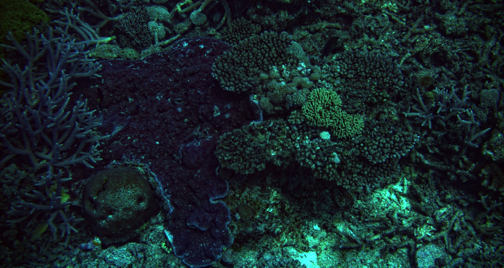}
  \end{minipage} 
       \begin{minipage}[t]{0.19\linewidth} 
    \centering 
    \includegraphics[width=1.32in, height=0.72in]{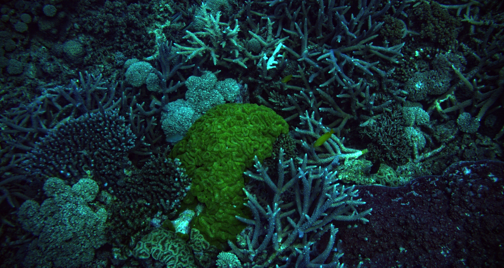}
  \end{minipage} 
       \begin{minipage}[t]{0.19\linewidth} 
    \centering 
    \includegraphics[width=1.32in, height=0.72in]{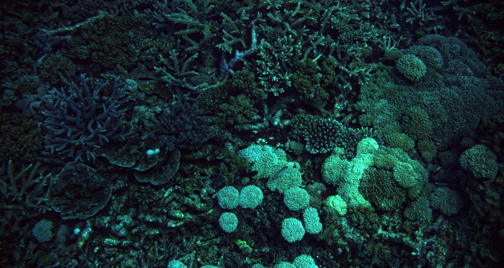}
  \end{minipage} 
         \begin{minipage}[t]{0.19\linewidth} 
    \centering 
    \includegraphics[width=1.32in, height=0.72in]{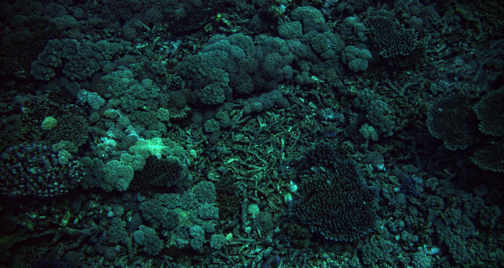}
  \end{minipage} 
  \\   
\begin{minipage}[t]{0.02\linewidth} 
    \centering
    \rotatebox{90}{ \;\; \;  \small IBLA}
              \vspace{-200mm}
      \end{minipage} 
     \begin{minipage}[t]{0.19\linewidth} 
    \centering 
    \includegraphics[width=1.32in, height=0.72in]{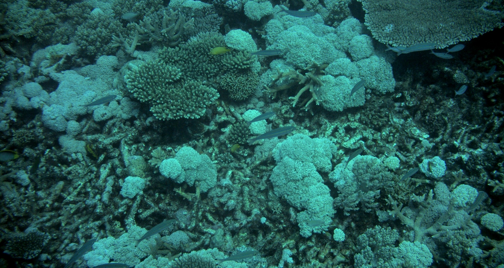}
  \end{minipage} 
       \begin{minipage}[t]{0.19\linewidth} 
    \centering 
    \includegraphics[width=1.32in, height=0.72in]{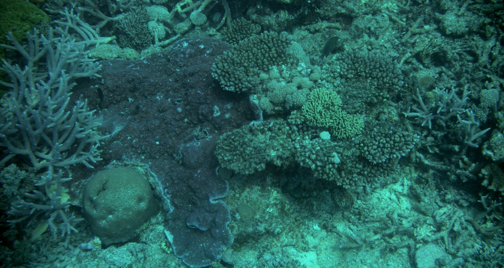}
  \end{minipage} 
       \begin{minipage}[t]{0.19\linewidth} 
    \centering 
    \includegraphics[width=1.32in, height=0.72in]{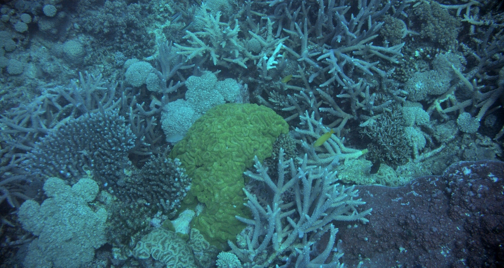}
  \end{minipage} 
       \begin{minipage}[t]{0.19\linewidth} 
    \centering 
    \includegraphics[width=1.32in, height=0.72in]{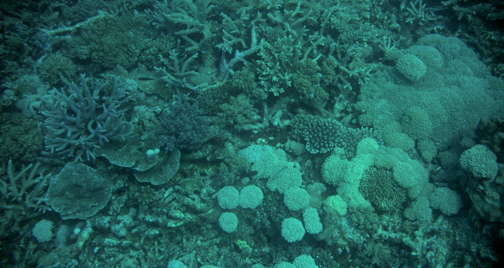}
  \end{minipage} 
         \begin{minipage}[t]{0.19\linewidth} 
    \centering 
    \includegraphics[width=1.32in, height=0.72in]{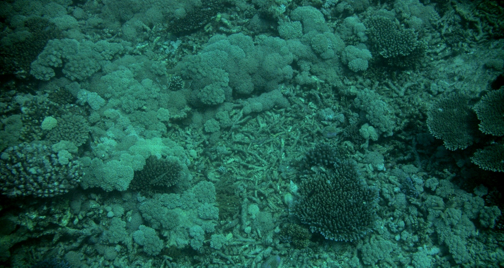}
  \end{minipage} 
  \\
\begin{minipage}[t]{0.02\linewidth} 
    \centering
    \rotatebox{90}{ \;\;  \small Haze-line}
              \vspace{-200mm}
      \end{minipage} 
     \begin{minipage}[t]{0.19\linewidth} 
    \centering 
    \includegraphics[width=1.32in, height=0.72in]{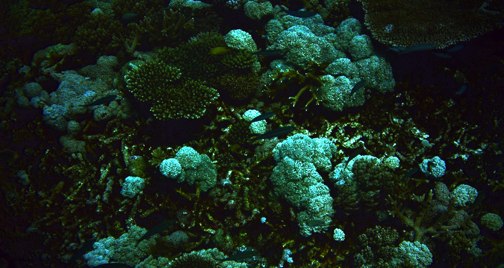}
  \end{minipage} 
       \begin{minipage}[t]{0.19\linewidth} 
    \centering 
    \includegraphics[width=1.32in, height=0.72in]{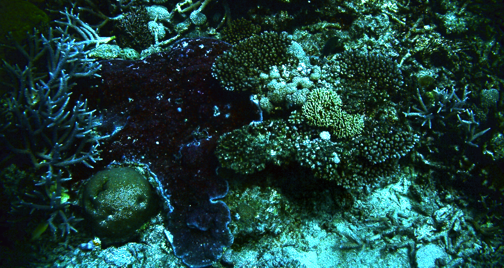}
  \end{minipage} 
       \begin{minipage}[t]{0.19\linewidth} 
    \centering 
    \includegraphics[width=1.32in, height=0.72in]{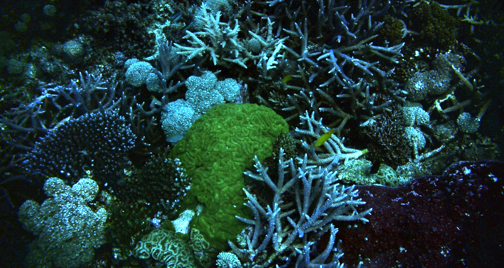}
  \end{minipage} 
       \begin{minipage}[t]{0.19\linewidth} 
    \centering 
    \includegraphics[width=1.32in, height=0.72in]{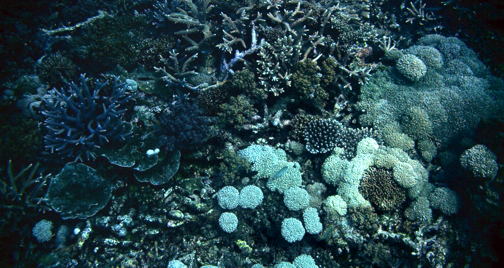}
  \end{minipage} 
         \begin{minipage}[t]{0.19\linewidth} 
    \centering 
    \includegraphics[width=1.32in, height=0.72in]{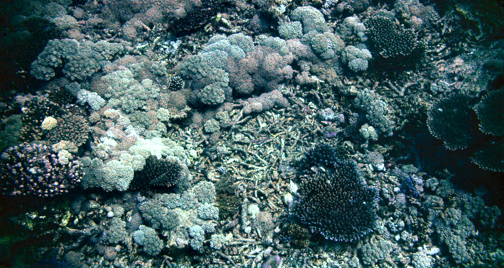}
  \end{minipage} 
  \\     
       \begin{minipage}[t]{0.02\linewidth} 
    \centering
    \rotatebox{90}{ \;\; \;  \small CUT}
      \vspace{-20mm}
      \end{minipage}
     \begin{minipage}[t]{0.19\linewidth} 
    \centering 
    \includegraphics[width=1.32in, height=0.72in]{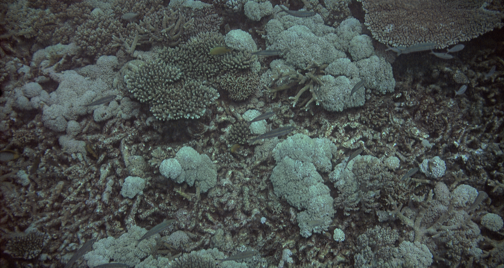}
  \end{minipage} 
       \begin{minipage}[t]{0.19\linewidth} 
    \centering 
    \includegraphics[width=1.32in, height=0.72in]{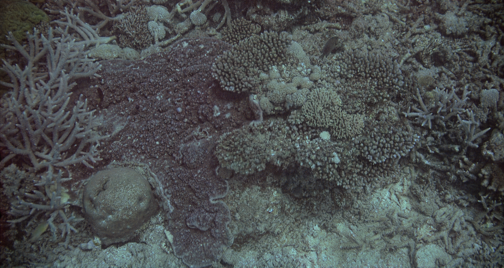}
  \end{minipage} 
       \begin{minipage}[t]{0.19\linewidth} 
    \centering 
    \includegraphics[width=1.32in, height=0.72in]{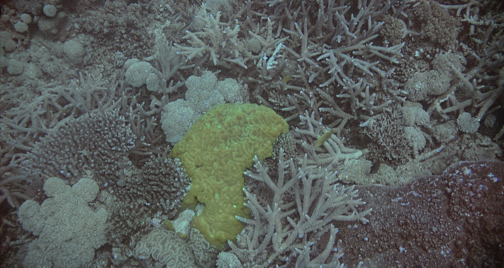}
  \end{minipage} 
       \begin{minipage}[t]{0.19\linewidth} 
    \centering 
    \includegraphics[width=1.32in, height=0.72in]{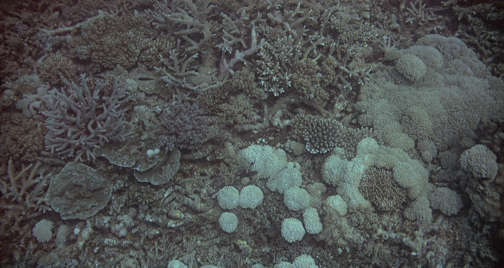}
  \end{minipage} 
         \begin{minipage}[t]{0.19\linewidth} 
    \centering 
    \includegraphics[width=1.32in, height=0.72in]{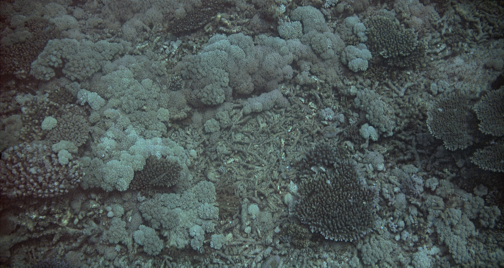}
  \end{minipage} 
  \\ 
     \begin{minipage}[t]{0.02\linewidth} 
    \centering
    \rotatebox{90}{\,\small CycleGAN}
              \vspace{-200mm}
      \end{minipage} 
     \begin{minipage}[t]{0.19\linewidth} 
    \centering 
    \includegraphics[width=1.32in, height=0.72in]{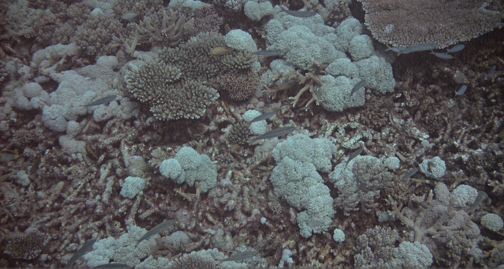}
  \end{minipage} 
       \begin{minipage}[t]{0.19\linewidth} 
    \centering 
    \includegraphics[width=1.32in, height=0.72in]{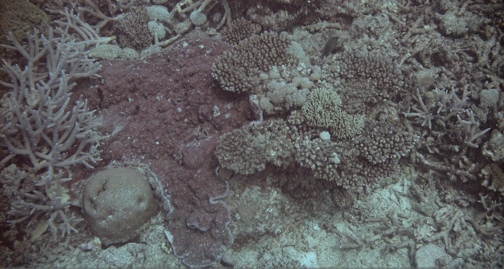}
  \end{minipage} 
       \begin{minipage}[t]{0.19\linewidth} 
    \centering 
    \includegraphics[width=1.32in, height=0.72in]{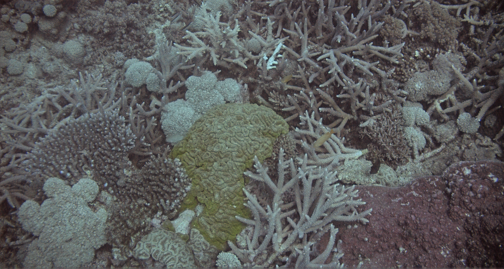}
  \end{minipage} 
       \begin{minipage}[t]{0.19\linewidth} 
    \centering 
    \includegraphics[width=1.32in, height=0.72in]{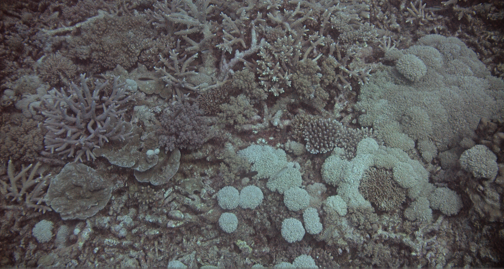}
  \end{minipage} 
         \begin{minipage}[t]{0.19\linewidth} 
    \centering 
    \includegraphics[width=1.32in, height=0.72in]{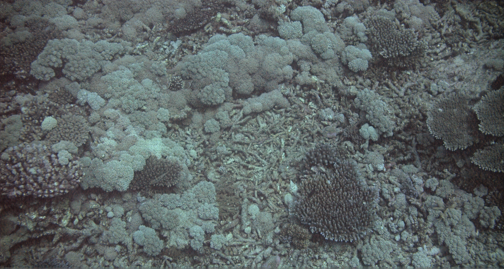}
  \end{minipage} 
  \\   
       \begin{minipage}[t]{0.02\linewidth} 
    \centering
    \rotatebox{90}{ \;  \small DCLGAN}
              \vspace{-200mm}
      \vspace{0.72mm}
      \end{minipage} 
     \begin{minipage}[t]{0.19\linewidth} 
    \centering 
    \includegraphics[width=1.32in, height=0.72in]{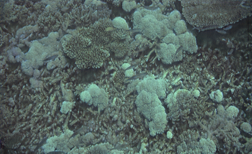}
  \end{minipage} 
       \begin{minipage}[t]{0.19\linewidth} 
    \centering 
    \includegraphics[width=1.32in, height=0.72in]{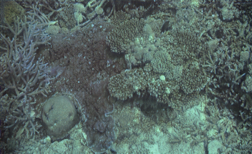}
  \end{minipage} 
       \begin{minipage}[t]{0.19\linewidth} 
    \centering 
    \includegraphics[width=1.32in, height=0.72in]{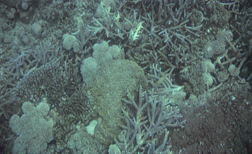}
  \end{minipage} 
       \begin{minipage}[t]{0.19\linewidth} 
    \centering 
    \includegraphics[width=1.32in, height=0.72in]{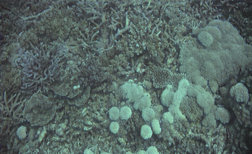}
  \end{minipage} 
         \begin{minipage}[t]{0.19\linewidth} 
    \centering 
    \includegraphics[width=1.32in, height=0.72in]{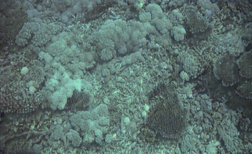}
  \end{minipage} 
  \\
\begin{minipage}[t]{0.02\linewidth} 
    \centering
    \rotatebox{90}{ \;\;   \small UWCNN}
              \vspace{-200mm}
      \end{minipage} 
     \begin{minipage}[t]{0.19\linewidth} 
    \centering 
    \includegraphics[width=1.32in, height=0.72in]{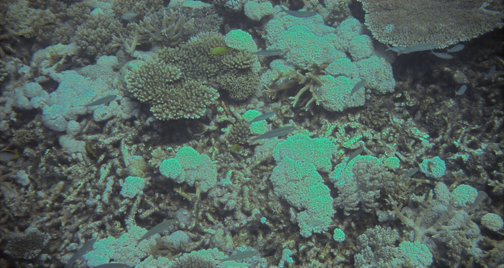}
  \end{minipage} 
       \begin{minipage}[t]{0.19\linewidth} 
    \centering 
    \includegraphics[width=1.32in, height=0.72in]{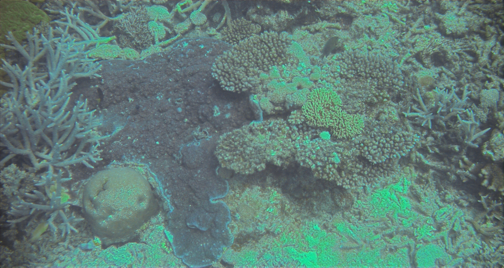}
  \end{minipage} 
       \begin{minipage}[t]{0.19\linewidth} 
    \centering 
    \includegraphics[width=1.32in, height=0.72in]{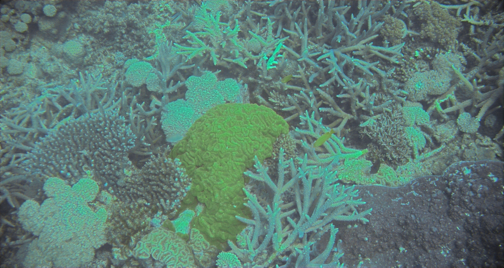}
  \end{minipage} 
       \begin{minipage}[t]{0.19\linewidth} 
    \centering 
    \includegraphics[width=1.32in, height=0.72in]{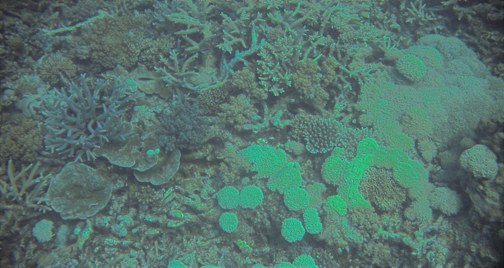}
  \end{minipage} 
         \begin{minipage}[t]{0.19\linewidth} 
    \centering 
    \includegraphics[width=1.32in, height=0.72in]{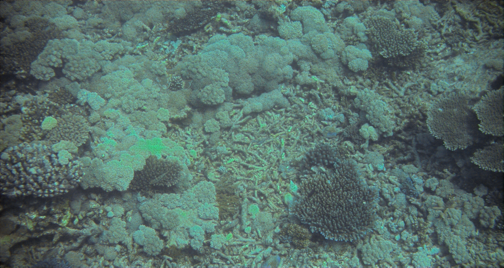}
  \end{minipage} 
  \\
 \begin{minipage}[t]{0.02\linewidth} 
    \centering
    \rotatebox{90}{\,\small CWR (ours)}
              \vspace{-200mm}
      \end{minipage} 
     \begin{minipage}[t]{0.19\linewidth} 
    \centering 
    \includegraphics[width=1.32in, height=0.72in]{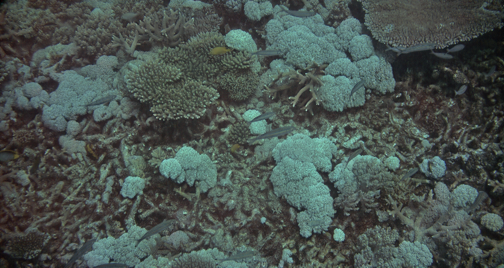}
  \end{minipage} 
       \begin{minipage}[t]{0.19\linewidth} 
    \centering 
    \includegraphics[width=1.32in, height=0.72in]{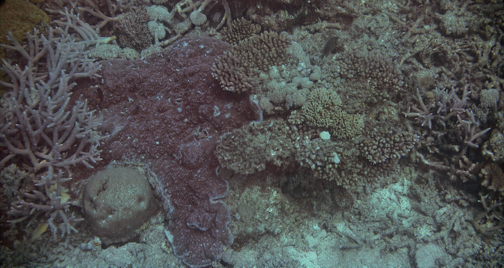}
  \end{minipage} 
       \begin{minipage}[t]{0.19\linewidth} 
    \centering 
    \includegraphics[width=1.32in, height=0.72in]{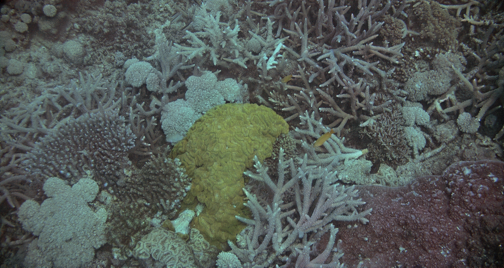}
  \end{minipage} 
       \begin{minipage}[t]{0.19\linewidth} 
    \centering 
    \includegraphics[width=1.32in, height=0.72in]{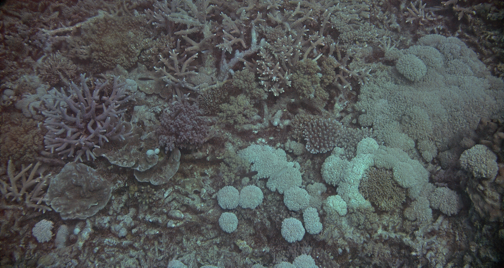}
  \end{minipage} 
         \begin{minipage}[t]{0.19\linewidth} 
    \centering 
    \includegraphics[width=1.32in, height=0.72in]{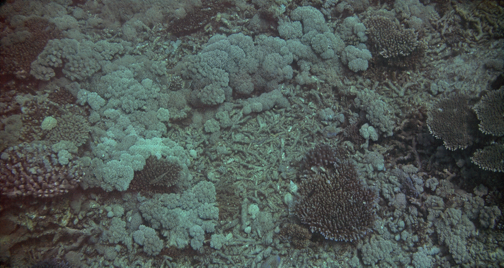}
  \end{minipage} 
  \\
   \begin{minipage}[t]{0.02\linewidth} 
    \centering
    \rotatebox{90}{ \;  \small Reference}
      \end{minipage} 
     \begin{minipage}[t]{0.19\linewidth} 
    \centering 
    \includegraphics[width=1.32in, height=0.72in]{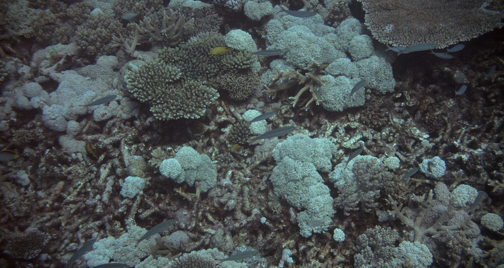}
  \end{minipage} 
       \begin{minipage}[t]{0.19\linewidth} 
    \centering 
    \includegraphics[width=1.32in, height=0.72in]{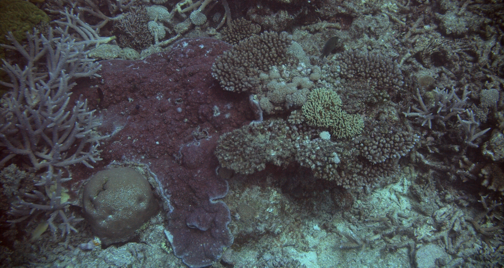}
  \end{minipage} 
       \begin{minipage}[t]{0.19\linewidth} 
    \centering 
    \includegraphics[width=1.32in, height=0.72in]{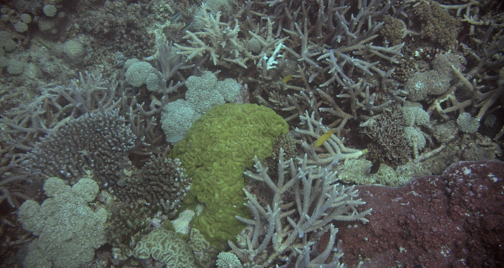}
  \end{minipage} 
       \begin{minipage}[t]{0.19\linewidth} 
    \centering 
    \includegraphics[width=1.32in, height=0.72in]{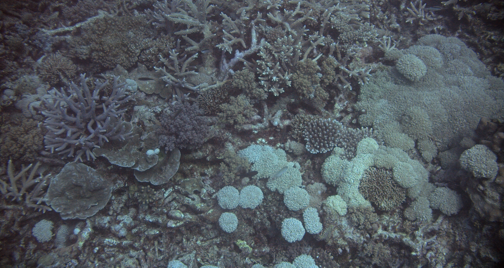}
  \end{minipage} 
         \begin{minipage}[t]{0.19\linewidth} 
    \centering 
    \includegraphics[width=1.32in, height=0.72in]{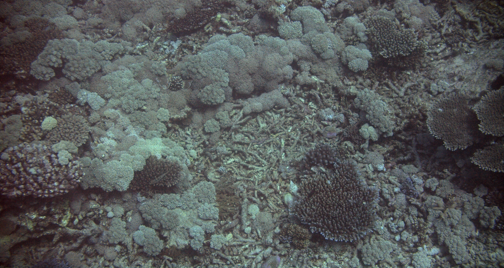}
  \end{minipage} 
   \caption{Qualitative results on the test set of HICRD, where all examples are randomly selected from the test set. We compare CWR to other underwater image restoration baselines. Conventional restoration methods fail to remove the green-bluish tone in underwater images. CWR shows visual satisfactory results without content and structure loss.}
 \label{fig:result1}
\end{figure*}

%% file: tables/compare.tex
\begin{table*}[!htbp]
  \centering
  \fontsize{11.5}{3}\selectfont
    \begin{tabular}{lccccccc}
    \toprule
    \textbf{Category} &\textbf{Method} &\textbf{MSE$\downarrow$} &\textbf{PSNR$\uparrow$} &\textbf{SSIM$\uparrow$} &\textbf{UIQM$\uparrow$} &\textbf{FID$\downarrow$}&\textbf{Speed$\downarrow$}   \cr
    \midrule
    \cmidrule(lr){1-8} 
    \multirow{6}{*}{CE}
    &Histogram~\cite{li2016hist}  & 2408.8    &14.44 & 0.618 & 5.27  
    &69.15 &10 \cr  
    &Retinex~\cite{fu2014retinex}    &1227.2  &17.36&0.722&\textcolor{red}{5.43}&71.90&5 \cr  
    &Fusion~\cite{ancuti2017color}   &1238.6  &17.53&0.783&\textcolor{blue}{5.33}&58.57&85\cr
    \cmidrule(lr){1-8} 
    \multirow{10}{*}{CR}
    &UDCP~\cite{drews2013transmission} &3159.9&13.31&0.489&4.99&38.03&67\cr  
    &DCP~\cite{he2010single}         &2548.2  &14.27&0.534&4.49&37.52&168\cr  
    &IBLA~\cite{peng2017underwater}  &803.9 &19.42&0.459&3.63&23.06&141\cr 
    &Haze-line~\cite{berman2020underwaterpami}  &2305.6&14.69&0.427&4.71&53.67&192\cr    
    \cmidrule(lr){1-8}
    \multirow{12}{*}{LR}
    &CUT~\cite{park2020contrastive}  &\textcolor{blue}{170.27}&\textcolor{blue}{26.30}&\textcolor{blue}{0.796}&5.26&22.35&46\cr
    &CycleGAN~\cite{CycleGAN2017}    &448.16    &21.81&0.591&5.27&\textcolor{red}{16.74}&46\cr
    &DCLGAN~\cite{han2021dcl}  &443.82&21.92&0.735&4.93&24.44&46\cr 
    &UWCNN~\cite{li2020underwater}   &775.8  &20.20&0.754&4.18&33.43&55\cr
    &CWR (ours) &\textcolor{red}{127.2}&\textcolor{red}{26.88}&\textcolor{red}{0.834}&5.25&\textcolor{blue}{18.20}&1.5/46\cr 
    \bottomrule
    \end{tabular}
     \caption{Comparisons to baselines on HICRD dataset on common evaluation metrics. Category shows the category of different methods, where L denotes Learning-based, R denotes Restoration, C denotes Conventional, and E denotes Enhancement. We show five quantitative metrics for all methods and the evaluation running speed. CWR is in the first place (in \textcolor{red}{red}) for MSE, PNSR, and SSIM while the second place (in \textcolor{blue}{blue}) for FID. Based on quantitative measurements, CWR produces restored images with higher quality. Speed refers to the evaluation speed per image in seconds. CWR also runs faster than all conventional restoration methods, and on par to other learning-based restoration methods. The speed of CWR with a Tesla P100 GPU is about 1.5 seconds per image while 46 seconds per image on an Intel(R) Core(TM) i5-6500 CPU @ 3.20GHZ.}
     \label{tab:3compare}
\end{table*}

%% file: figures/result_enhancement2.tex
   \begin{figure*}[!htbp]
   \begin{minipage}[t]{0.03\linewidth} 
  \text{}
  \end{minipage} 
    \begin{minipage}[t]{0.23\linewidth} 
    \centering 
    \text{Example 2}
  \end{minipage}
      \begin{minipage}[t]{0.12\linewidth} 
    \centering 
    \text{Zoomed 1}
  \end{minipage}
      \begin{minipage}[t]{0.12\linewidth} 
    \centering 
    \text{Zoomed 2}
  \end{minipage}
  \begin{minipage}[t]{0.23\linewidth} 
    \centering 
    \text{Example 3}
  \end{minipage}   
 \begin{minipage}[t]{0.12\linewidth} 
    \centering 
    \text{Zoomed 3}
  \end{minipage}
      \begin{minipage}[t]{0.12\linewidth} 
    \centering 
    \text{Zoomed 4}
  \end{minipage}
  \\
\begin{minipage}[t]{0.03\linewidth} 
    \centering
    \rotatebox{90}{ \quad  Input}
   \vspace{-20mm}
      \end{minipage} 
     \begin{minipage}[t]{0.23\linewidth} 
    \centering 
    \includegraphics[width=1.55in, height=0.8in]{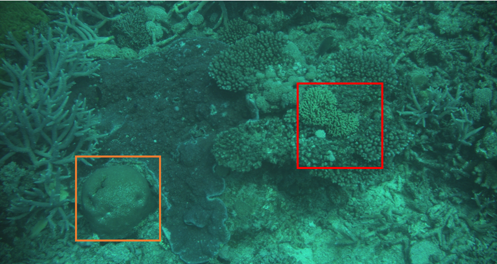}
  \end{minipage} 
     \begin{minipage}[t]{0.12\linewidth} 
    \centering 
    \includegraphics[width=0.8in, height=0.8in]{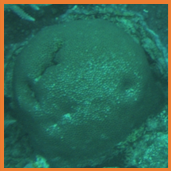}
  \end{minipage} 
     \begin{minipage}[t]{0.12\linewidth} 
    \centering 
    \includegraphics[width=0.8in, height=0.8in]{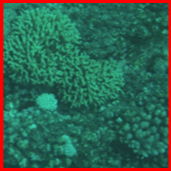}
  \end{minipage} 
     \begin{minipage}[t]{0.23\linewidth} 
    \centering 
    \includegraphics[width=1.55in, height=0.8in]{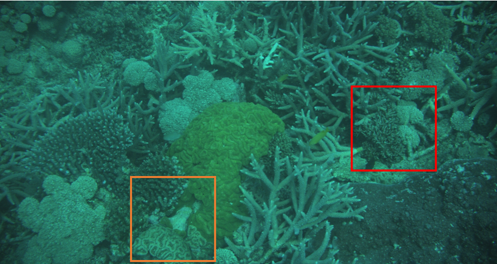}
  \end{minipage} 
     \begin{minipage}[t]{0.12\linewidth} 
    \centering 
    \includegraphics[width=0.8in, height=0.8in]{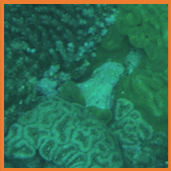}
  \end{minipage} 
     \begin{minipage}[t]{0.12\linewidth} 
    \centering 
    \includegraphics[width=0.8in, height=0.8in]{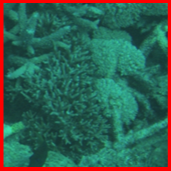}
  \end{minipage} 
  \\
  \begin{minipage}[t]{0.03\linewidth} 
    \centering
    \rotatebox{90}{\;\;Histogram}
   \vspace{-20mm}
      \end{minipage} 
     \begin{minipage}[t]{0.23\linewidth} 
    \centering 
    \includegraphics[width=1.55in, height=0.8in]{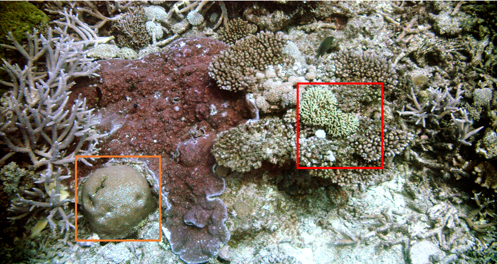}
  \end{minipage} 
     \begin{minipage}[t]{0.12\linewidth} 
    \centering 
    \includegraphics[width=0.8in, height=0.8in]{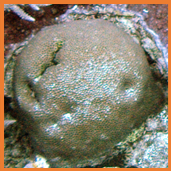}
  \end{minipage} 
     \begin{minipage}[t]{0.12\linewidth} 
    \centering 
    \includegraphics[width=0.8in, height=0.8in]{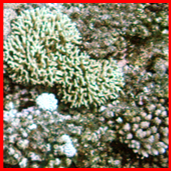}
  \end{minipage} 
     \begin{minipage}[t]{0.23\linewidth} 
    \centering 
    \includegraphics[width=1.55in, height=0.8in]{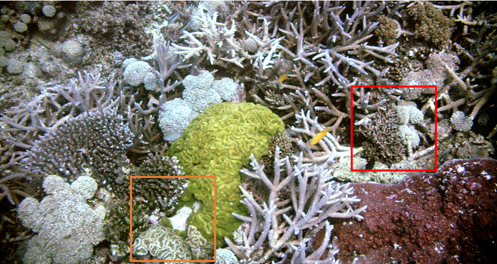}
  \end{minipage} 
     \begin{minipage}[t]{0.12\linewidth} 
    \centering 
    \includegraphics[width=0.8in, height=0.8in]{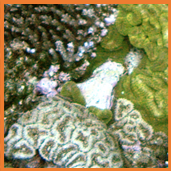}
  \end{minipage} 
     \begin{minipage}[t]{0.12\linewidth} 
    \centering 
    \includegraphics[width=0.8in, height=0.8in]{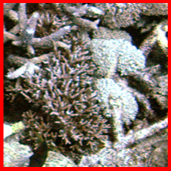}
  \end{minipage} 
  \\
    \begin{minipage}[t]{0.03\linewidth} 
    \centering
    \rotatebox{90}{ \quad Retinex}
   \vspace{-20mm}
      \end{minipage} 
     \begin{minipage}[t]{0.23\linewidth} 
    \centering 
    \includegraphics[width=1.55in, height=0.8in]{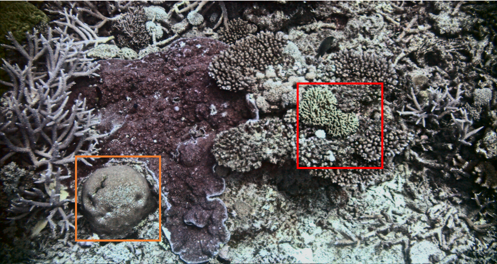}
  \end{minipage} 
     \begin{minipage}[t]{0.12\linewidth} 
    \centering 
    \includegraphics[width=0.8in, height=0.8in]{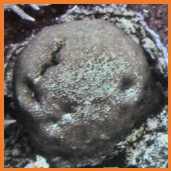}
  \end{minipage} 
     \begin{minipage}[t]{0.12\linewidth} 
    \centering 
    \includegraphics[width=0.8in, height=0.8in]{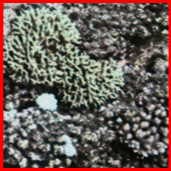}
  \end{minipage} 
     \begin{minipage}[t]{0.23\linewidth} 
    \centering 
    \includegraphics[width=1.55in, height=0.8in]{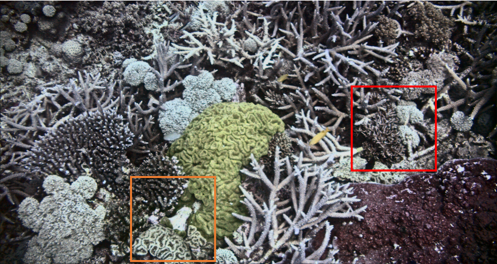}
  \end{minipage} 
     \begin{minipage}[t]{0.12\linewidth} 
    \centering 
    \includegraphics[width=0.8in, height=0.8in]{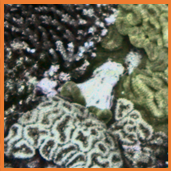}
  \end{minipage} 
     \begin{minipage}[t]{0.12\linewidth} 
    \centering 
    \includegraphics[width=0.8in, height=0.8in]{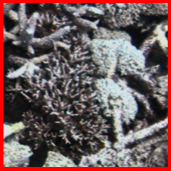}
  \end{minipage} 
  \\
      \begin{minipage}[t]{0.03\linewidth} 
    \centering
    \rotatebox{90}{\quad \;Fusion}
   \vspace{-20mm}
      \end{minipage} 
     \begin{minipage}[t]{0.23\linewidth} 
    \centering 
    \includegraphics[width=1.55in, height=0.8in]{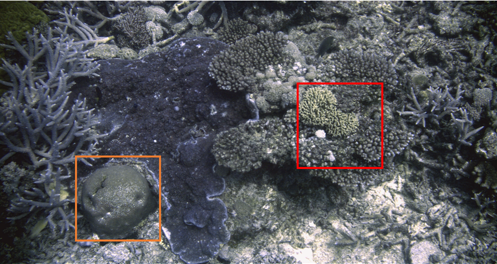}
  \end{minipage} 
     \begin{minipage}[t]{0.12\linewidth} 
    \centering 
    \includegraphics[width=0.8in, height=0.8in]{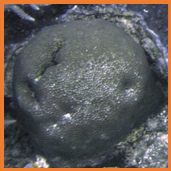}
  \end{minipage} 
     \begin{minipage}[t]{0.12\linewidth} 
    \centering 
    \includegraphics[width=0.8in, height=0.8in]{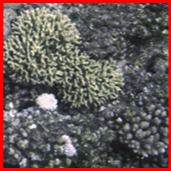}
  \end{minipage} 
     \begin{minipage}[t]{0.23\linewidth} 
    \centering 
    \includegraphics[width=1.55in, height=0.8in]{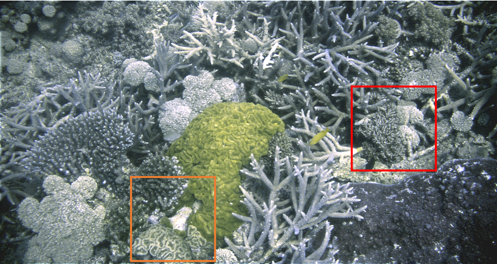}
  \end{minipage} 
     \begin{minipage}[t]{0.12\linewidth} 
    \centering 
    \includegraphics[width=0.8in, height=0.8in]{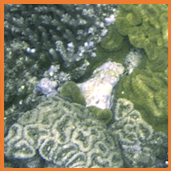}
  \end{minipage} 
     \begin{minipage}[t]{0.12\linewidth} 
    \centering 
    \includegraphics[width=0.8in, height=0.8in]{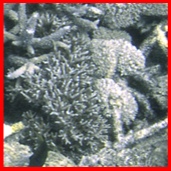}
  \end{minipage} 
  \\
        \begin{minipage}[t]{0.03\linewidth} 
    \centering
    \rotatebox{90}{\; \;Reference}
   \vspace{-20mm}
      \end{minipage} 
     \begin{minipage}[t]{0.23\linewidth} 
    \centering 
    \includegraphics[width=1.55in, height=0.8in]{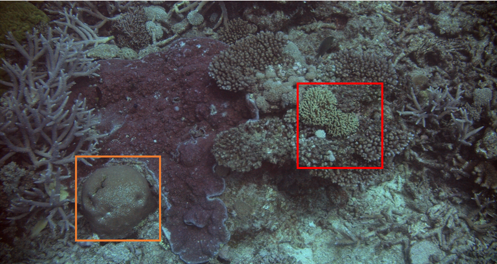}
  \end{minipage} 
     \begin{minipage}[t]{0.12\linewidth} 
    \centering 
    \includegraphics[width=0.8in, height=0.8in]{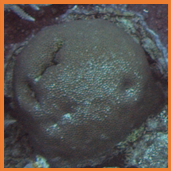}
  \end{minipage} 
     \begin{minipage}[t]{0.12\linewidth} 
    \centering 
    \includegraphics[width=0.8in, height=0.8in]{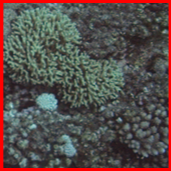}
  \end{minipage} 
     \begin{minipage}[t]{0.23\linewidth} 
    \centering 
    \includegraphics[width=1.55in, height=0.8in]{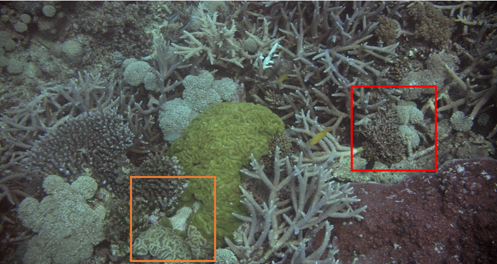}
  \end{minipage} 
     \begin{minipage}[t]{0.12\linewidth} 
    \centering 
    \includegraphics[width=0.8in, height=0.8in]{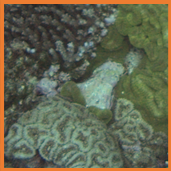}
  \end{minipage} 
     \begin{minipage}[t]{0.12\linewidth} 
    \centering 
    \includegraphics[width=0.8in, height=0.8in]{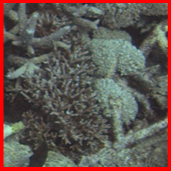}
  \end{minipage} 
   
   \caption{Qualitative results on the test set of HICRD for underwater image enhancement methods. Example 2 and Example 3 are identical to figure~\ref{fig:result1}. Enhancement methods usually produce aesthetically pleasing results, however, enhancement methods sometimes over enhance the images, adding bright color to the stone and coral reef, which against the true color of the objects. The zoomed section presents the details of color and structure.}
 \label{fig:result2}
\end{figure*}

%% file: bare_jrnl.bbl
\begin{thebibliography}{10}
\providecommand{\url}[1]{#1}
\csname url@samestyle\endcsname
\providecommand{\newblock}{\relax}
\providecommand{\bibinfo}[2]{#2}
\providecommand{\BIBentrySTDinterwordspacing}{\spaceskip=0pt\relax}
\providecommand{\BIBentryALTinterwordstretchfactor}{4}
\providecommand{\BIBentryALTinterwordspacing}{\spaceskip=\fontdimen2\font plus
\BIBentryALTinterwordstretchfactor\fontdimen3\font minus
  \fontdimen4\font\relax}
\providecommand{\BIBforeignlanguage}[2]{{%
\expandafter\ifx\csname l@#1\endcsname\relax
\typeout{** WARNING: IEEEtran.bst: No hyphenation pattern has been}%
\typeout{** loaded for the language `#1'. Using the pattern for}%
\typeout{** the default language instead.}%
\else
\language=\csname l@#1\endcsname
\fi
#2}}
\providecommand{\BIBdecl}{\relax}
\BIBdecl

\bibitem{reggiannini2021use}
M.~Reggiannini and D.~Moroni, ``The use of saliency in underwater computer
  vision: A review,'' \emph{Remote Sensing}, vol.~13, no.~1, p.~22, 2021.

\bibitem{williams2014exploiting}
D.~P. Williams and E.~Fakiris, ``Exploiting environmental information for
  improved underwater target classification in sonar imagery,'' \emph{IEEE
  Transactions on Geoscience and Remote Sensing}, vol.~52, no.~10, pp.
  6284--6297, 2014.

\bibitem{ludeno2018microwave}
G.~Ludeno, L.~Capozzoli, E.~Rizzo, F.~Soldovieri, and I.~Catapano, ``A
  microwave tomography strategy for underwater imaging via ground penetrating
  radar,'' \emph{Remote Sensing}, vol.~10, no.~9, p. 1410, 2018.

\bibitem{fei2014contributions}
T.~Fei, D.~Kraus, and A.~M. Zoubir, ``Contributions to automatic target
  recognition systems for underwater mine classification,'' \emph{IEEE
  Transactions on Geoscience and Remote Sensing}, vol.~53, no.~1, pp. 505--518,
  2014.

\bibitem{carlevaris2010initial}
N.~Carlevaris-Bianco, A.~Mohan, and R.~M. Eustice, ``Initial results in
  underwater single image dehazing,'' in \emph{Oceans 2010 Mts/IEEE
  Seattle}.\hskip 1em plus 0.5em minus 0.4em\relax IEEE, 2010, pp. 1--8.

\bibitem{Akkaynak_2018_CVPR}
D.~Akkaynak and T.~Treibitz, ``A revised underwater image formation model,'' in
  \emph{Proceedings of the IEEE Conference on Computer Vision and Pattern
  Recognition (CVPR)}, June 2018, pp. 6723--6732.

\bibitem{yuan2020underwater}
J.~Yuan, W.~Cao, Z.~Cai, and B.~Su, ``An underwater image vision enhancement
  algorithm based on contour bougie morphology,'' \emph{IEEE Transactions on
  Geoscience and Remote Sensing}, 2020.

\bibitem{he2010single}
K.~He, J.~Sun, and X.~Tang, ``Single image haze removal using dark channel
  prior,'' \emph{IEEE transactions on pattern analysis and machine
  intelligence}, vol.~33, no.~12, pp. 2341--2353, 2010.

\bibitem{drews2013transmission}
P.~Drews, E.~Nascimento, F.~Moraes, S.~Botelho, and M.~Campos, ``Transmission
  estimation in underwater single images,'' in \emph{Proceedings of the IEEE
  international conference on computer vision workshops}, 2013, pp. 825--830.

\bibitem{galdran2015automatic}
A.~Galdran, D.~Pardo, A.~Pic{\'o}n, and A.~Alvarez-Gila, ``Automatic
  red-channel underwater image restoration,'' \emph{Journal of Visual
  Communication and Image Representation}, vol.~26, pp. 132--145, 2015.

\bibitem{chiang2011underwater}
J.~Y. Chiang and Y.-C. Chen, ``Underwater image enhancement by wavelength
  compensation and dehazing,'' \emph{IEEE transactions on image processing},
  vol.~21, no.~4, pp. 1756--1769, 2011.

\bibitem{lu2015contrast}
H.~Lu, Y.~Li, L.~Zhang, and S.~Serikawa, ``Contrast enhancement for images in
  turbid water,'' \emph{JOSA A}, vol.~32, no.~5, pp. 886--893, 2015.

\bibitem{peng2017underwater}
Y.-T. Peng and P.~C. Cosman, ``Underwater image restoration based on image
  blurriness and light absorption,'' \emph{IEEE transactions on image
  processing}, vol.~26, no.~4, pp. 1579--1594, 2017.

\bibitem{jerlov1976marine}
N.~G. Jerlov, \emph{Marine optics}.\hskip 1em plus 0.5em minus 0.4em\relax
  Elsevier, 1976.

\bibitem{berman2020underwaterpami}
D.~Berman, D.~Levy, S.~Avidan, and T.~Treibitz, ``Underwater single image color
  restoration using haze-lines and a new quantitative dataset,'' \emph{IEEE
  transactions on pattern analysis and machine intelligence}, 2020.

\bibitem{cao2018underwaterestorationr}
K.~Cao, Y.-T. Peng, and P.~C. Cosman, ``Underwater image restoration using deep
  networks to estimate background light and scene depth,'' in \emph{2018 IEEE
  Southwest Symposium on Image Analysis and Interpretation (SSIAI)}.\hskip 1em
  plus 0.5em minus 0.4em\relax IEEE, 2018, pp. 1--4.

\bibitem{barbosa2018visualrestoration}
W.~V. Barbosa, H.~G. Amaral, T.~L. Rocha, and E.~R. Nascimento,
  ``Visual-quality-driven learning for underwater vision enhancement,'' in
  \emph{2018 25th IEEE International Conference on Image Processing
  (ICIP)}.\hskip 1em plus 0.5em minus 0.4em\relax IEEE, 2018, pp. 3933--3937.

\bibitem{cai2016dehazenet}
B.~Cai, X.~Xu, K.~Jia, C.~Qing, and D.~Tao, ``Dehazenet: An end-to-end system
  for single image haze removal,'' \emph{IEEE Transactions on Image
  Processing}, vol.~25, no.~11, pp. 5187--5198, 2016.

\bibitem{hou2018jointrestoration}
M.~Hou, R.~Liu, X.~Fan, and Z.~Luo, ``Joint residual learning for underwater
  image enhancement,'' in \emph{2018 25th IEEE International Conference on
  Image Processing (ICIP)}.\hskip 1em plus 0.5em minus 0.4em\relax IEEE, 2018,
  pp. 4043--4047.

\bibitem{li2020underwater}
C.~Li, S.~Anwar, and F.~Porikli, ``Underwater scene prior inspired deep
  underwater image and video enhancement,'' \emph{Pattern Recognition},
  vol.~98, p. 107038, 2020.

\bibitem{huang2017densely}
G.~Huang, Z.~Liu, L.~Van Der~Maaten, and K.~Q. Weinberger, ``Densely connected
  convolutional networks,'' in \emph{Proceedings of the IEEE conference on
  computer vision and pattern recognition}, 2017, pp. 4700--4708.

\bibitem{li2019underwater}
C.~Li, C.~Guo, W.~Ren, R.~Cong, J.~Hou, S.~Kwong, and D.~Tao, ``An underwater
  image enhancement benchmark dataset and beyond,'' \emph{IEEE Transactions on
  Image Processing}, vol.~29, pp. 4376--4389, 2019.

\bibitem{duarte2016dataset}
A.~Duarte, F.~Codevilla, J.~D.~O. Gaya, and S.~S. Botelho, ``A dataset to
  evaluate underwater image restoration methods,'' in \emph{OCEANS
  2016-Shanghai}.\hskip 1em plus 0.5em minus 0.4em\relax IEEE, 2016, pp. 1--6.

\bibitem{fabbri2018enhancing}
C.~Fabbri, M.~J. Islam, and J.~Sattar, ``Enhancing underwater imagery using
  generative adversarial networks,'' in \emph{2018 IEEE International
  Conference on Robotics and Automation (ICRA)}.\hskip 1em plus 0.5em minus
  0.4em\relax IEEE, 2018, pp. 7159--7165.

\bibitem{islam2020fast}
M.~J. Islam, Y.~Xia, and J.~Sattar, ``Fast underwater image enhancement for
  improved visual perception,'' \emph{IEEE Robotics and Automation Letters},
  vol.~5, no.~2, pp. 3227--3234, 2020.

\bibitem{wang2019underwater}
K.~Wang, Y.~Hu, J.~Chen, X.~Wu, X.~Zhao, and Y.~Li, ``Underwater image
  restoration based on a parallel convolutional neural network,'' \emph{Remote
  sensing}, vol.~11, no.~13, p. 1591, 2019.

\bibitem{li2018emerging}
C.~Li, J.~Guo, and C.~Guo, ``Emerging from water: Underwater image color
  correction based on weakly supervised color transfer,'' \emph{IEEE Signal
  processing letters}, vol.~25, no.~3, pp. 323--327, 2018.

\bibitem{Silberman:ECCV12NYU}
P.~K. Nathan~Silberman, Derek~Hoiem and R.~Fergus, ``Indoor segmentation and
  support inference from rgbd images,'' in \emph{ECCV}, 2012.

\bibitem{Akkaynak_2019_CVPR}
D.~Akkaynak and T.~Treibitz, ``Sea-thru: A method for removing water from
  underwater images,'' in \emph{Proceedings of the IEEE/CVF Conference on
  Computer Vision and Pattern Recognition (CVPR)}, June 2019, pp. 1682--1691.

\bibitem{anwar2020survey}
S.~Anwar and C.~Li, ``Diving deeper into underwater image enhancement: A
  survey,'' \emph{Signal Processing: Image Communication}, vol.~89, p. 115978,
  2020.

\bibitem{he2020momentum}
K.~He, H.~Fan, Y.~Wu, S.~Xie, and R.~Girshick, ``Momentum contrast for
  unsupervised visual representation learning,'' in \emph{IEEE Conference on
  Computer Vision and Pattern Recognition (CVPR)}, 2020, pp. 9729--9738.

\bibitem{chen2020simple}
T.~Chen, S.~Kornblith, M.~Norouzi, and G.~Hinton, ``A simple framework for
  contrastive learning of visual representations,'' in \emph{International
  Conference on Machine Learning (ICML)}, 2020, pp. 1597--1607.

\bibitem{goodfellow2014generative}
I.~Goodfellow, J.~Pouget-Abadie, M.~Mirza, B.~Xu, D.~Warde-Farley, S.~Ozair,
  A.~Courville, and Y.~Bengio, ``Generative adversarial nets,'' in
  \emph{Advances in neural information processing systems (NIPS)}, 2014, pp.
  2672--2680.

\bibitem{han2021cwr}
J.~Han, M.~Shoeiby, T.~Malthus, E.~Botha, J.~Anstee, S.~Anwar, R.~Wei,
  L.~Petersson, and M.~A. Armin, ``Single underwater image restoration by
  contrastive learning,'' in \emph{IEEE International Geoscience and Remote
  Sensing Symposium (IGARSS)}, 2021.

\bibitem{schonberg2007bioeroding}
C.~H. Sch{\"o}nberg and R.~Suwa, ``Why bioeroding sponges may be better hosts
  for symbiotic dinoflagellates than many corals,'' \emph{Porifera Research:
  Biodiversity, Innovation and Sustainability. Publ Museu Nac Rio de Janeiro},
  pp. 569--580, 2007.

\bibitem{reef_check2018}
J.~Salmond, J.~Passenger, E.~Kovacs, C.~Roelfsema, and D.~Stetner, \emph{Reef
  Check Australia 2018 Heron Island Reef Health Report.}\hskip 1em plus 0.5em
  minus 0.4em\relax Reef Check Foundation Ltd., 2018.

\bibitem{boss_twardowski}
E.~Boss, M.~Twardowski, D.~McKee, I.~Cetinić, and W.~Slade, \emph{Beam
  Transmission and Attenuation Coefficients: Instruments, Characterization,
  Field Measurements and Data Analysis Protocols.}, 2nd~ed., ser. IOCCG Ocean
  Optics and Biogeochemistry Protocols for Satellite Ocean Colour Sensor
  Validation.\hskip 1em plus 0.5em minus 0.4em\relax IOCCG, 2019.

\bibitem{https://doi.org/10.1002/2014JC010205}
K.~Oubelkheir, P.~W. Ford, L.~A. Clementson, N.~Cherukuru, G.~Fry, and A.~D.~L.
  Steven, ``Impact of an extreme flood event on optical and biogeochemical
  properties in a subtropical coastal periurban embayment (eastern
  australia),'' \emph{Journal of Geophysical Research: Oceans}, vol. 119,
  no.~9, pp. 6024--6045, 2014.

\bibitem{mannino_novak}
A.~Mannino, M.~G. Novak, N.~B. Nelson, M.~Belz, J.-F. Berthon, N.~V. Blough,
  E.~Boss, A.~Brichaud, J.~Chaves, C.~Del~Castillo, and et~al.,
  \emph{Measurement protocol of absorption by chromophoric dissolved organic
  matter (CDOM) and other dissolved materials}, 1st~ed., ser. IOCCG Ocean
  Optics and Biogeochemistry Protocols for Satellite Ocean Colour Sensor
  Validation.\hskip 1em plus 0.5em minus 0.4em\relax IOCCG, 2019.

\bibitem{Austin1981}
\BIBentryALTinterwordspacing
R.~W. Austin and T.~J. Petzold, \emph{The Determination of the Diffuse
  Attenuation Coefficient of Sea Water Using the Coastal Zone Color
  Scanner}.\hskip 1em plus 0.5em minus 0.4em\relax Boston, MA: Springer US,
  1981, pp. 239--256. [Online]. Available:
  \url{https://doi.org/10.1007/978-1-4613-3315-9_29}
\BIBentrySTDinterwordspacing

\bibitem{Simon:13}
\BIBentryALTinterwordspacing
A.~Simon and P.~Shanmugam, ``A new model for the vertical spectral diffuse
  attenuation coefficient of downwelling irradiance in turbid coastal waters:
  validation with in situ measurements,'' \emph{Opt. Express}, vol.~21, no.~24,
  pp. 30\,082--30\,106, Dec 2013. [Online]. Available:
  \url{http://www.opticsexpress.org/abstract.cfm?URI=oe-21-24-30082}
\BIBentrySTDinterwordspacing

\bibitem{serikawa2014underwater}
S.~Serikawa and H.~Lu, ``Underwater image dehazing using joint trilateral
  filter,'' \emph{Computers \& Electrical Engineering}, vol.~40, no.~1, pp.
  41--50, 2014.

\bibitem{park2020contrastive}
T.~Park, A.~A. Efros, R.~Zhang, and J.-Y. Zhu, ``Contrastive learning for
  unpaired image-to-image translation,'' in \emph{European Conference on
  Computer Vision}, 2020, pp. 319--345.

\bibitem{CycleGAN2017}
J.-Y. Zhu, T.~Park, P.~Isola, and A.~A. Efros, ``Unpaired image-to-image
  translation using cycle-consistent adversarial networks,'' in
  \emph{Proceedings of the IEEE international conference on computer vision},
  2017, pp. 2223--2232.

\bibitem{he2016deep}
K.~He, X.~Zhang, S.~Ren, and J.~Sun, ``Deep residual learning for image
  recognition,'' in \emph{IEEE Conference on Computer vision and pattern
  recognitio (CVPR)}, 2016, pp. 770--778.

\bibitem{ulyanov2016instance}
D.~Ulyanov, A.~Vedaldi, and V.~Lempitsky, ``Instance normalization: The missing
  ingredient for fast stylization,'' \emph{arXiv preprint arXiv:1607.08022},
  2016.

\bibitem{isola2017image}
P.~Isola, J.-Y. Zhu, T.~Zhou, and A.~A. Efros, ``Image-to-image translation
  with conditional adversarial networks,'' in \emph{IEEE Conference on Computer
  Vision and Pattern Recognition (CVPR)}, 2017, pp. 1125--1134.

\bibitem{miyato2018spectral}
T.~Miyato, T.~Kataoka, M.~Koyama, and Y.~Yoshida, ``Spectral normalization for
  generative adversarial networks,'' \emph{arXiv preprint arXiv:1802.05957},
  2018.

\bibitem{han2021dcl}
J.~Han, M.~Shoeiby, L.~Petersson, and M.~A. Armin, ``Dual contrastive learning
  for unsupervised image-to-image translation,'' in \emph{Proceedings of the
  IEEE/CVF Conference on Computer Vision and Pattern Recognition Workshops},
  2021.

\bibitem{li2016hist}
C.-Y. Li, J.-C. Guo, R.-M. Cong, Y.-W. Pang, and B.~Wang, ``Underwater image
  enhancement by dehazing with minimum information loss and histogram
  distribution prior,'' \emph{IEEE Transactions on Image Processing}, vol.~25,
  no.~12, pp. 5664--5677, 2016.

\bibitem{fu2014retinex}
X.~Fu, P.~Zhuang, Y.~Huang, Y.~Liao, X.-P. Zhang, and X.~Ding, ``A
  retinex-based enhancing approach for single underwater image,'' in
  \emph{International Conference on Image Processing}, 2014, pp. 4572--4576.

\bibitem{ancuti2017color}
C.~O. Ancuti, C.~Ancuti, C.~De~Vleeschouwer, and P.~Bekaert, ``Color balance
  and fusion for underwater image enhancement,'' \emph{IEEE Transactions on
  image processing}, vol.~27, no.~1, pp. 379--393, 2017.

\bibitem{kingma2014adam}
D.~P. Kingma and J.~Ba, ``Adam: A method for stochastic optimization,''
  \emph{International Conference on Learning Representations (ICLR)}, 2014.

\bibitem{ssim2014}
{Zhou Wang}, A.~C. {Bovik}, H.~R. {Sheikh}, and E.~P. {Simoncelli}, ``Image
  quality assessment: from error visibility to structural similarity,''
  \emph{IEEE Transactions on Image Processing}, vol.~13, no.~4, pp. 600--612,
  2004.

\bibitem{panetta2015human}
K.~Panetta, C.~Gao, and S.~Agaian, ``Human-visual-system-inspired underwater
  image quality measures,'' \emph{IEEE Journal of Oceanic Engineering},
  vol.~41, no.~3, pp. 541--551, 2015.

\bibitem{TTUR}
M.~Heusel, H.~Ramsauer, T.~Unterthiner, B.~Nessler, and S.~Hochreiter, ``Gans
  trained by a two time-scale update rule converge to a local nash
  equilibrium,'' in \emph{Advances in neural information processing systems},
  2017.

\bibitem{mangeruga2018guidelines}
M.~Mangeruga, F.~Bruno, M.~Cozza, P.~Agrafiotis, and D.~Skarlatos, ``Guidelines
  for underwater image enhancement based on benchmarking of different
  methods,'' \emph{Remote Sensing}, vol.~10, no.~10, p. 1652, 2018.

\bibitem{berman2017diving}
D.~Berman, T.~Treibitz, and S.~Avidan, ``Diving into haze-lines: Color
  restoration of underwater images,'' in \emph{Proc. British Machine Vision
  Conference (BMVC)}, vol.~1, no.~2, 2017.

\bibitem{Akkaynak_2017_CVPR}
D.~Akkaynak, T.~Treibitz, T.~Shlesinger, Y.~Loya, R.~Tamir, and D.~Iluz, ``What
  is the space of attenuation coefficients in underwater computer vision?'' in
  \emph{Proceedings of the IEEE Conference on Computer Vision and Pattern
  Recognition (CVPR)}, July 2017, pp. 4931--4940.

\bibitem{liu2020real}
R.~Liu, X.~Fan, M.~Zhu, M.~Hou, and Z.~Luo, ``Real-world underwater
  enhancement: Challenges, benchmarks, and solutions under natural light,''
  \emph{IEEE Transactions on Circuits and Systems for Video Technology},
  vol.~30, no.~12, pp. 4861--4875, 2020.

\bibitem{yi2018instantaneous}
D.~H. Yi, Z.~Gong, J.~M. Jech, P.~Ratilal, and N.~C. Makris, ``Instantaneous 3d
  continental-shelf scale imaging of oceanic fish by multi-spectral resonance
  sensing reveals group behavior during spawning migration,'' \emph{Remote
  Sensing}, vol.~10, no.~1, p. 108, 2018.

\bibitem{fu2020underwaterhyper}
X.~Fu, X.~Shang, X.~Sun, H.~Yu, M.~Song, and C.-I. Chang, ``Underwater
  hyperspectral target detection with band selection,'' \emph{Remote Sensing},
  vol.~12, no.~7, p. 1056, 2020.

\bibitem{mogstad2019shallow}
A.~A. Mogstad, G.~Johnsen, and M.~Ludvigsen, ``Shallow-water habitat mapping
  using underwater hyperspectral imaging from an unmanned surface vehicle: a
  pilot study,'' \emph{Remote Sensing}, vol.~11, no.~6, p. 685, 2019.

\bibitem{dumke2018underwater}
I.~Dumke, M.~Ludvigsen, S.~L. Ellefmo, F.~S{\o}reide, G.~Johnsen, and B.~J.
  Murton, ``Underwater hyperspectral imaging using a stationary platform in the
  trans-atlantic geotraverse hydrothermal field,'' \emph{IEEE Transactions on
  Geoscience and Remote Sensing}, vol.~57, no.~5, pp. 2947--2962, 2018.

\bibitem{guo2016model}
Y.~Guo, H.~Song, H.~Liu, H.~Wei, P.~Yang, S.~Zhan, H.~Wang, H.~Huang, N.~Liao,
  Q.~Mu \emph{et~al.}, ``Model-based restoration of underwater spectral images
  captured with narrowband filters,'' \emph{Optics express}, vol.~24, no.~12,
  pp. 13\,101--13\,120, 2016.

\end{thebibliography}
